\documentstyle[aas2pp4]{article}
\slugcomment{Submitted to {\it The Astrophysical Journal}\/ June 1996}

\lefthead{Hamilton et al.}
\righthead{UV Lines in SN 1006}
\def\bib{\reference{}}
\newcommand{\rmn}{\rm}
\newcommand{\be}{\begin{equation}}
\newcommand{\ee}{\end{equation}}
\newcommand{\ba}{\begin{eqnarray}}
\newcommand{\ea}{\end{eqnarray}}
\newcommand{\nn}{\nonumber \\}

\newcommand{\cm}{{\rmn cm}}
\newcommand{\dd}{d}		% ApJ
\newcommand{\erg}{{\rmn erg}}
\newcommand{\gm}{{\rmn gm}}
\newcommand{\s}{{\rmn s}}
\newcommand{\yr}{{\rmn yr}}
\newcommand{\kms}{{\rmn km}\,{\rmn s}^{-1}}
\newcommand{\kpc}{{\rmn kpc}}
\newcommand{\Msun}{{\rmn M}_{\sun}}
\newcommand{\Fe}{{\rmn Fe}}
\newcommand{\FeII}{{\rmn Fe II}}	% MNRAS
\newcommand{\Ni}{{\rmn Ni}}
\newcommand{\Si}{{\rmn Si}}
\newcommand{\SiII}{{\rmn Si II}}	% ApJ
\newcommand{\SiIII}{{\rmn Si III}}	% ApJ
\newcommand{\SiIV}{{\rmn Si IV}}	% ApJ

\begin{document}

%--------------------
% ApJ version
\title{Interpretation of UV Absorption Lines in SN1006\footnote{
Based on observations made with the NASA/ESA Hubble Space Telescope,
obtained at the Space Telescope Science Institute, which is operated by
the Association of Universities for Research in Astronomy, Inc.,
under NASA contract NAS 5-26555.}}

\author{A. J. S. Hamilton}
\affil{JILA and Dept.\ of Astrophysical, Planetary and Atmospheric Sciences,
Box 440, University of Colorado, Boulder, CO 80309; ajsh@dark.colorado.edu}

\author{R. A. Fesen}
\affil{Dept.\ of Physics \& Astronomy, 6127 Wilder Lab., Dartmouth College,
Hanover, NH 03755; fesen@oak.dartmouth.edu}

\author{C.-C. Wu}
\affil{Computer Sciences Corporation,
STScI, 3700 San Martin Drive, Baltimore, MD 21218; wu@stsci.edu}

\author{D. M. Crenshaw}
\affil{Computer Sciences Corporation,
LASP,Code 681, NASA-GSFC, Greenbelt, MD 20771; hrsmike@hrs.gsfc.nasa.gov}

\and

\author{C. L. Sarazin}
\affil{Dept.\ Astronomy, U.\ Virginia,
Box 3818, Charlotesville, VA 22903-0818; cls7i@coma.astro.virginia.edu}

\vspace*{3em}{\topsep 0pt\center\large
Submitted to {\it The Astrophysical Journal}\/ June 1996
\endcenter}
%--------------------

%%--------------------
%% MNRAS version
%\maketitle
%%--------------------

\begin{abstract}
We present a theoretical interpretation of the broad silicon
and iron ultraviolet absorption features observed with
the Hubble Space Telescope
in the spectrum of the Schweizer-Middleditch star behind the remnant of
Supernova 1006.
These features are caused by supernova ejecta in SN1006.
We propose that the redshifted \ion{Si}{2} 1260\,\AA\ feature consists of
both unshocked and shocked \ion{Si}{2}.
%accounting for the curious profile of this line.
The sharp red edge of the line at $7070 \,\kms$ indicates the position
of the reverse shock,
while its Gaussian blue edge reveals shocked Si with a mean velocity
of $5050 \,\kms$ and a dispersion of $1240 \,\kms$,
implying a reverse shock velocity of $2860 \,\kms$.
The measured velocities satisfy the energy jump condition for a strong shock,
provided that all the shock energy goes into ions,
with little or no collisionless heating of electrons.
%This \ion{Si}{2} 1260\,\AA\ line provides a remarkable
%demonstration of the existence of a strong shock under highly collisionless
%conditions.
The line profiles of the \ion{Si}{3} and \ion{Si}{4} absorption features
indicate that they arise mostly from shocked Si.
The total mass of shocked and unshocked Si inferred from the
\ion{Si}{2}, \ion{Si}{3} and \ion{Si}{4} profiles
is $M_\Si = 0.25 \pm 0.01 \,\Msun$ on the assumption of spherical symmetry.
Unshocked Si extends upwards from $5600 \,\kms$.
Although there appears to be some Fe mixed with the Si at lower velocities
$\la 7070 \,\kms$,
the absence of \ion{Fe}{2} absorption with the same profile as the
shocked \ion{Si}{2} suggests little Fe mixed with Si
at higher (before being shocked) velocities.
The column density of shocked \ion{Si}{2} is close to that expected
for \ion{Si}{2} undergoing steady state collisional ionization
behind the reverse shock,
provided that the electron to \ion{Si}{2} ratio is low,
from which we infer that most of the shocked Si is likely to be
of a fairly high degree of purity, unmixed with other elements.
We propose that the ambient interstellar density on the far side of SN1006
is anomalously low compared to the density around the rest of the remnant.
This would simultaneously explain
the high velocity of the redshifted Si absorption,
the absence of blueshifted Si absorption,
and the low density of the absorbing Si compared to the high Si
density required to produce the observed Si x-ray line emission.
We have reanalyzed the \ion{Fe}{2} absorption lines,
concluding that the earlier evidence for high velocity
blueshifted \ion{Fe}{2} extending to $\sim - 8000 \,\kms$ is not compelling.
We interpret the blue edge on the \ion{Fe}{2} profiles at $- 4200 \,\kms$
as the position of the reverse shock on the near side of SN1006.
The mass of \ion{Fe}{2} inferred from the red edge of the
\ion{Fe}{2} profile is
$M_\FeII = 0.029 \pm 0.004 \,\Msun$ up to $7070 \,\kms$,
if spherical symmetry is assumed.
The low ionization state of unshocked Si inferred from
our analysis of the silicon features, \ion{Si}{2}/Si $= 0.92 \pm 0.07$,
suggests a correspondingly low ionization state of unshocked iron,
with \ion{Fe}{2}/Fe = $0.66^{+ 0.29}_{- 0.22}$.
If this is correct,
then the total mass of Fe up to $7070 \,\kms$ is
$M_\Fe = 0.044^{+ 0.022}_{- 0.013} \,\Msun$
with a $3 \sigma$ upper limit of $M_\Fe < 0.16 \,\Msun$.
Such a low ionization state and mass of iron is consistent
with the recent observation of \ion{Fe}{3} 1123\,\AA\ with HUT,
indicating \ion{Fe}{3}/\ion{Fe}{2} $= 1.1 \pm 0.9$,
but conflicts with the expected presence of several tenths of a solar mass of
iron in this suspected Type~Ia remnant.
However,
the inference from the present HST data is too indirect,
and the HUT data are too noisy,
to rule out a large mass of iron.
Re-observation of the \ion{Fe}{3} 1123\,\AA\ line at higher signal to noise
ratio with FUSE will be important in determining the degree of ionization
and hence mass of iron in SN1006.

\end{abstract}

%--------------------
% ApJ version
%--------------------

%%--------------------
%% MNRAS version
%\begin{keywords}
%supernova remnants -- ISM: SN1006 -- ultraviolet: ISM
%\end{keywords}
%
%%\clearpage
%%--------------------

\section{Introduction}
\label{intro}

In two previous papers, we (Wu et al.\ 1993, 1996, hereafter WCFHS93, WCHFLS96)
described HST FOS observations of the UV spectrum of the
SM star (Schweizer \& Middleditch 1980) which lies
behind and close to the projected center of the remnant of SN1006.
In the present paper we offer a theoretical interpretation of
the broad silicon and iron UV absorption features observed with HST.
These features are almost certainly caused by supernova ejecta in SN1006,
as originally proposed by Wu et al.\ (1983),
who first observed the features with IUE.

Detailed theoretical analysis of the \ion{Fe}{2} features observed
with IUE has been presented previously by
Hamilton \& Fesen (1988, hereafter HF88).
The main purpose of that paper was to try to explain the apparent
conflict between the low
$\approx 0.015 \,\Msun$
mass of \ion{Fe}{2} inferred from the IUE observations of SN1006
(Fesen et al.\ 1988)
with the expected presence of several tenths of a solar mass of iron
(H\"{o}flich \& Khokhlov 1996)
in this suspected Type~Ia remnant
(Minkowski 1966; Schaefer 1996).
HF88 demonstrated that ambient UV starlight and UV and x-ray emission
from reverse-shocked ejecta could photoionize unshocked iron
mainly to \ion{Fe}{3}, \ion{Fe}{4}, and \ion{Fe}{5},
resolving the conflict.

Recently
Blair, Long \& Raymond (1996) used the Hopkins Ultraviolet Telescope (HUT)
to measure \ion{Fe}{3} 1123\,\AA\ absorption in the spectrum of the SM star.
They found \ion{Fe}{3}/\ion{Fe}{2} $= 1.1 \pm 0.9$,
which neither confirms, nor excludes,
the ratio \ion{Fe}{3}/\ion{Fe}{2} $= 2.6$ predicted by HF88.

The HST spectra, particularly the silicon features,
prove to be a rich source of information
beyond the reach of IUE's capabilities.
In the first half of this paper,
Section~\ref{silicon},
we analyze the Si absorption features.
We find (\S\ref{red}) that
the profile of the redshifted \ion{Si}{2} 1260\,\AA\ feature,
with its sharp red edge and Gaussian blue edge,
appears to be attributable to the presence
of both unshocked and shocked silicon.
We then develop a chain of inferences,
first about the reverse shock (\S\ref{jump}) and collisionless heating
(\S\ref{collisionless}),
then about the column density and mass (\S\S\ref{mass}, \ref{Simass}),
purity (\S\ref{purity}), and ionization state (\S\ref{SiIII+IV})
of the silicon.
We argue (\S\ref{blue}) that the ambient interstellar density on the far of
SN1006 is anomalously low compared to density around the rest of the remnant,
which explains the high velocity of the redshifted Si,
the absence of corresponding blueshifted Si (\S\ref{blueion}),
and some other observational puzzles.

In the second half of the paper,
Section~\ref{iron},
we discuss the broad \ion{Fe}{2} absorption features.
WCFHS93 reported blueshifted \ion{Fe}{2} absorption up to $\sim - 8000 \,\kms$.
Finding the presence of such high velocity blueshifted absorption
difficult to understand in the light of other observational evidence,
we detail a reanalysis of the \ion{Fe}{2} features in \S\ref{reanalysis}.
We conclude (\S\ref{fe2sec}) that the evidence for high velocity blueshifted
absorption is not compelling,
and we propose (\S\ref{nearside}) that the sharp blue edge on the
\ion{Fe}{2} profiles at $- 4200 \,\kms$ represents the position of the
reverse shock on the near side.
In the remainder of the Section,
\S\S\ref{shockedFe}-\ref{where},
we address the issue of the ionization state and mass of the iron.

We attempt in this paper to construct a consistent theoretical picture,
but there remain some discrepancies,
and we highlight these in Section~\ref{worries}.
Section~\ref{summary} summarizes the conclusions.

\section{Silicon}
\label{silicon}

There are three possibilities for the origin of the broad
Si absorption features in the spectrum of the SM star,
if it is accepted that these features arise from the remnant of SN1006.

The first possibility is that the absorption arises from cool, dense,
fast moving knots of ejecta, as suggested by Fesen et al.\ (1988)
and further discussed by Fesen \& Hamilton (1988).
However, there has been no significant change in the features over 12 years,
from the original detection of the features with IUE in 1982 (Wu et al.\ 1983),
through their re-observation with IUE in 1986 and 1988 (Fesen \& Hamilton 1988),
up to the FOS observation with HST in 1994 (WCHFLS96).
Furthermore,
the relative strengths of the redshifted
\ion{Si}{2} 1260\,\AA, 1304\,\AA, and 1527\,\AA\ features
are approximately proportional to their oscillator strengths
times wavelengths, indicating that the lines are not saturated.
This constancy in time and lack of saturation
argues against the absorption features being caused by small, dense knots.
We do not consider this hypothesis further in this paper.

A second possibility is that the Si absorption is from shocked ejecta
in which the collisional ionization timescale is so long that
the observed low ionization species
\ion{Si}{2}, \ion{Si}{3}, and \ion{Si}{4} can survive.
At first sight,
the high $\sim 5000\,\kms$ velocity of the observed absorption
makes this possibility seem unlikely,
because shocked ejecta should be moving no faster than the velocity
of gas behind the interstellar shock,
which in the NW sector of the remnant can be inferred
from the $2310\,\kms$ FWHM of the Balmer broad line emission
to be
$1800$-$2400\,\kms$
(this is $3/4$ of the shock velocity),
depending on the extent to which electron-ion equilibration takes place
(Kirshner, Winkler \& Chevalier 1987;
Long, Blair \& van den Bergh 1988;
Smith et al.\ 1991;
Raymond, Blair \& Long 1995).
However, below we will conclude that it is likely that
much, in fact most, of the Si absorption is from shocked ejecta,
and that the ISM surrounding SN1006 may be quite inhomogeneous.

A third possibility is that the Si absorption arises from
unshocked supernova ejecta freely expanding in SN1006,
which is consistent with the high velocity of the absorption.
The low ionization state of the Si,
predominantly \ion{Si}{2}, with some \ion{Si}{3} and \ion{Si}{4},
is at least qualitatively consistent with the expectations of models
in which the unshocked ejecta are photoionized by ambient UV starlight
and by UV and x-ray emission from shocked ejecta
(HF88).
Neutral Si is neither observed, e.g.\ at \ion{Si}{1} 1845\,\AA,
nor expected, since it should be quickly ($\sim 20 \, {\rmn yr}$)
photoionized by ambient UV starlight.
Recombination is negligible at the low densities here.
At the outset therefore, this possibility seems most likely,
and we pursue the idea further in the next subsection.

Fesen et al.\ (1988)
pointed out that the redshifted \ion{Si}{2} 1260\,\AA\ feature
(at $\sim 1280$\,\AA) in the IUE data
appeared to be somewhat too strong compared to the weaker redshifted
\ion{Si}{2} 1527\,\AA\ feature.
The discrepancy appears to be confirmed by the HST observations (WCHFLS96).
Fesen et al.\ proposed that some of the \ion{Si}{2} 1260\,\AA\ feature
may come from
\ion{S}{2} 1260, 1254, 1251\,\AA\
redshifted by $\approx 800 \,\kms$ relative to the \ion{Si}{2},
a possibility also addressed by WCHFLS96.
In the present paper we regard the possibility of any significant contribution
from \ion{S}{2} as unlikely,
notwithstanding the discrepancy between the \ion{Si}{2} profiles.
The oscillator strengths of the 
\ion{S}{2} 1260, 1254, 1251\,\AA\
lines are
$f = 0.01624$, 0.01088, 0.005453,
whose combined strength is only 1/30 of the oscillator strength
$f = 1.007$ of the \ion{Si}{2} 1260\,\AA\ line
(Morton 1991).
In models of Type~Ia supernovae, such as SN1006 is believed to be,
silicon and sulfur occur typically in the same region of space,
with a relative abundance of $\mbox{Si} : \mbox{S} \approx 2 : 1$
(e.g.\ Nomoto, Thielemann \& Yokoi 1984).
Thus \ion{S}{2} might be expected to contribute only $\sim 1/60$ of the
optical depth of \ion{Si}{2} in the 1260\,\AA\ feature,
assuming a similar ionization state of Si and S.
In the remainder of this paper we ignore any possible contribution
of \ion{S}{2} to the \ion{Si}{2} 1260\,\AA\ feature.
In this we follow Wu et al.'s (1983) original identification of
the 1280\,\AA\ absorption feature as redshifted Si.

\subsection{The redshifted \protect\ion{Si}{2} 1260\,\AA\ feature}
\label{red}

\begin{figure}[tb]
\epsfbox[170 262 415 525]{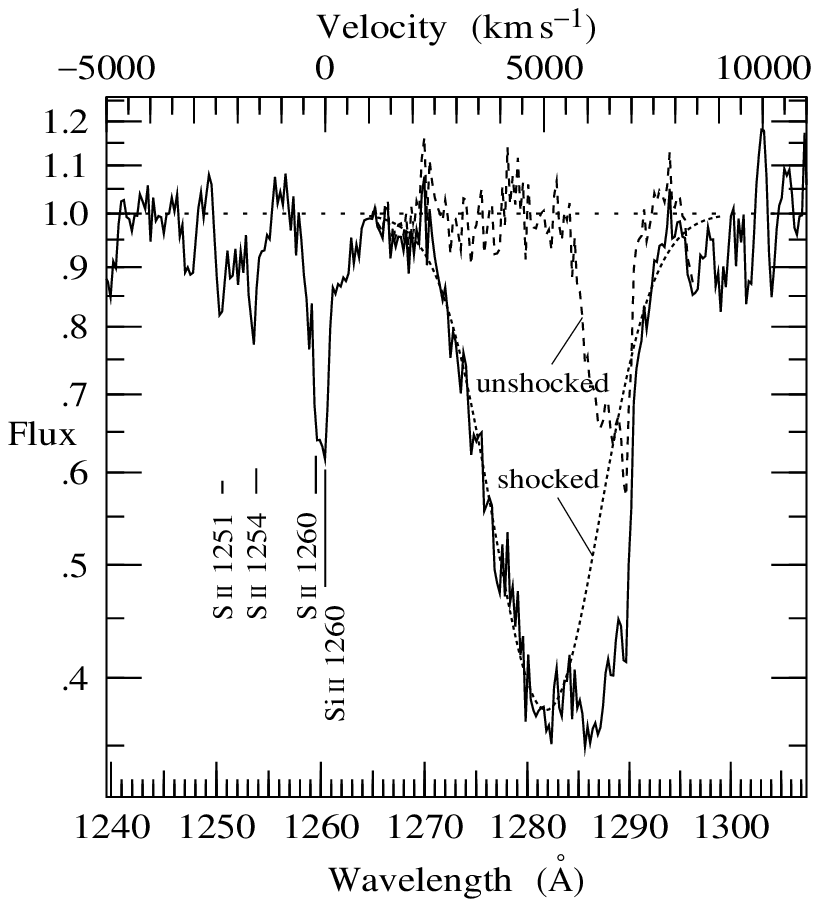}
%\vbox to94mm{\rule{0pt}{94mm}}
%\special{psfile=si1260.ps
%         angle=0 hoffset=-175 voffset=-265 vscale=100 hscale=100}
  \caption[1]{
HST G130H spectrum relative to the stellar continuum around
the redshifted \protect\ion{Si}{2} 1260.4221\,\AA\ feature,
showing the best fit Gaussian profile of shocked \protect\ion{Si}{2},
and the residual unshocked \protect\ion{Si}{2}.
Upper axis shows velocity in the rest frame of the \protect\ion{Si}{2} line.
Measured parameters of the feature are given in Table~\protect\ref{redtab}.
As elsewhere in this paper,
we assume a stellar continuum which is linear in $\log F$-$\log \lambda$,
and fit the continuum to ostensibly uncontaminated regions around the line.
The uncertainties in the parameters given in Table~\protect\ref{redtab}
include uncertainty from placement of the continuum.
The adopted stellar continuum is
$\log F =
\log ( 4.1 \times 10^{-14} \,\erg\,\s^{-1}\,\cm^{-2}\,{\rmn \AA}^{-1} )
- 2.3 \log (\lambda/1260\,{\rmn \AA})$.
\label{si1260}
}
\end{figure}

%--------------------
% ApJ version
\begin{deluxetable}{lr}
\tablewidth{0pt}
\tablecaption{Parameters measured from redshifted
\protect\ion{Si}{2} 1260\,\AA\ feature
\label{redtab}}
\tablehead{\colhead{Parameter} & \colhead{Value}}
\startdata
Expansion velocity into reverse shock & $7070 \pm 50 \,\kms$ \nl
Mean velocity of shocked \protect\ion{Si}{2} & $5050 \pm 60 \,\kms$ \nl
Dispersion of shocked \protect\ion{Si}{2} & $1240 \pm 40 \,\kms$ \nl
Reverse shock velocity & $2860 \pm 100 \,\kms$ \nl
Lower edge of unshocked \protect\ion{Si}{2} & $5600 \pm 100 \,\kms$ \nl
Preshock density of \protect\ion{Si}{2} &
$5.4 \pm 0.7 \times 10^{-5} \,\cm^{-3} $ \nl
Column density of shocked \protect\ion{Si}{2} &
$9.0 \pm 0.3 \times 10^{14} \,\cm^{-2}$ \nl
Column density of unshocked \protect\ion{Si}{2} &
$1.5 \pm 0.2 \times 10^{14} \,\cm^{-2}$ \nl
Column density of all \protect\ion{Si}{2} &
$10.5 \pm 0.1 \times 10^{14} \,\cm^{-2}$ \nl
Mass of shocked \protect\ion{Si}{2}
	& $0.127 \pm 0.006 \,\Msun$ \nl
Mass of unshocked \protect\ion{Si}{2} & $0.017 \pm 0.002 \,\Msun$ \nl
\enddata
\tablecomments{
Masses assume spherical symmetry.
For simplicity,
velocities and masses have not been adjusted for the small offset
of the SM star from the projected center of the remnant.
}
\end{deluxetable}
%--------------------

Given that SN1006 shows a well-developed interstellar blast wave
in both radio
(Reynolds \& Gilmore 1986, 1993)
and x-rays
(Koyama et al.\ 1995;
Willingale et al.\ 1996),
it is inevitable that a reverse shock must be propagating into
any unshocked ejecta.

If the observed UV Si absorption is from unshocked ejecta,
then there should be a sharp cutoff in the line profile
at the expansion velocity of the reverse shock,
because the shock should `instantaneously' decelerate the ejecta to
lower bulk velocity.
In fact the \ion{Si}{2} 1260\,\AA\ feature does show a steep red edge
at $7070 \pm 50 \,\kms$,
albeit with a possible tail to higher velocities.
Tentatively,
we take the presence of the steep red edge as evidence
that at least some of the \ion{Si}{2} is unshocked.

Once shocked, how long will \ion{Si}{2} last before being collisionally
ionized to higher levels?
The ionization timescale of \ion{Si}{2} entering the reverse shock
can be inferred from the optical depth of the
\ion{Si}{2} 1260\,\AA\ absorption just inside its steep red edge
at $7070\,\kms$,
which putatively represents the position of the reverse shock.
In freely expanding ejecta where radius equals velocity times age,
$r = v t$,
the column density per unit velocity $\dd N/\dd v$ of any species
is equal to the density $n = \dd N/\dd r$ times age $t$.
For freely expanding ejecta,
the optical depth $\tau$
in a line of wavelength $\lambda$ and oscillator strength $f$
is then proportional to $n t$:
\be
\label{tau}
	\tau
	= {\pi e^2 \over m_e c} f \lambda {\dd N \over \dd v}
	= {\pi e^2 \over m_e c} f \lambda n t
	\ .
\ee
The optical depth of the \ion{Si}{2} 1260.4221\,\AA\ line
($f = 1.007$, Morton 1991)
just inside its steep red edge is $\tau \approx 1$.
This implies, from equation (\ref{tau}),
a preshock \ion{Si}{2} density times age of
$n_\SiII^{\rm presh} t = 3.0 \times 10^6 \cm^{-3} \s$.
The postshock \ion{Si}{2} density times age would then be 4 times higher
for a strong shock,
$n_\SiII t = 1.2 \times 10^7 \cm^{-3} \s$.
At a collisional ionization rate for \ion{Si}{2} of
$\langle \sigma v \rangle_\SiII = 6 \times 10^{-8} \cm^3 \s^{-1}$
(Lennon et al.\ 1988;
see subsection \ref{purity} below for further discussion of this rate),
the ratio of the \ion{Si}{2} ionization timescale
$t_\SiII \equiv ( n_e \langle \sigma v \rangle_\SiII )^{-1}$
to the age of the remnant, $t$, is
(the following estimate is revised below, equation [\ref{tion'}])
\be
\label{tion}
	{t_\SiII \over t}
	= {1 \over n_e t \langle \sigma v \rangle_\SiII}
	= {n_\SiII \over n_e} {1 \over n_\SiII t \langle \sigma v \rangle_\SiII}
	= 1.4 {n_\SiII \over n_e}
	\ .
\ee
Since the ratio $n_e / n_\SiII$ of electron to \ion{Si}{2} density
in the postshock gas should be greater than but of order unity,
this estimate (\ref{tion})
indicates that the collisional timescale of \ion{Si}{2} is of
the order of the age of the remnant.
It follows that shocked \ion{Si}{2} is likely also to contribute to the
observed \ion{Si}{2} absorption.

%%--------------------
%% MNRAS version
%\begin{table}
%%\begin{table*}
%%\begin{minipage}{100mm}
%\caption{Parameters measured from redshifted \ion{Si}{2} 1260\,\AA\ feature
%\label{redtab}}
%%\begin{tabular}{@{}l@{}r}
%\begin{tabular}{@{}lr}
%Parameter & \multicolumn{1}{c}{Value} \\ [4pt]
%Expansion velocity into reverse shock & $7070 \pm 50 \,\kms$ \\
%Mean velocity of shocked \ion{Si}{2} & $5050 \pm 60 \,\kms$ \\
%Dispersion of shocked \ion{Si}{2} & $1240 \pm 40 \,\kms$ \\
%Reverse shock velocity & $2860 \pm 100 \,\kms$ \\
%Lower edge of unshocked \ion{Si}{2} & $5600 \pm 100 \,\kms$ \\
%Preshock density of \ion{Si}{2} & $5.4 \pm 0.7 \times 10^{-5} \,\cm^{-3}$ \\
%Column density of shocked \ion{Si}{2} &
%$9.0 \pm 0.3 \times 10^{14} \,\cm^{-2}$ \\
%Column density of unshocked \ion{Si}{2} &
%$1.5 \pm 0.2 \times 10^{14} \,\cm^{-2}$ \\
%Column density of all \ion{Si}{2} & $10.5 \pm 0.1 \times 10^{14} \,\cm^{-2}$ \\
%Mass of shocked \ion{Si}{2}
%	& $0.127 \pm 0.006 \,\Msun$ \\
%Mass of unshocked \ion{Si}{2} & $0.017 \pm 0.002 \,\Msun$ \\ [5pt]
%\end{tabular}
%Masses assume spherical symmetry.
%For simplicity,
%velocities and masses have not been adjusted for the small offset
%of the SM star from the projected center of the remnant.
%%\end{minipage}
%%\end{table*}
%\end{table}
%%--------------------

Shocked \ion{Si}{2} will be decelerated by the reverse shock to lower
velocities than the freely expanding unshocked ejecta.
Shocked \ion{Si}{2} should have a broad thermal profile,
unlike the unshocked \ion{Si}{2}.
Examining the redshifted \ion{Si}{2} 1260\,\AA\ profile,
we see that the blue edge extends down to about $+ 2500\,\kms$,
with a shape which looks Gaussian.
Fitting the blue edge to a Gaussian,
we find a best fit to a Gaussian centered at $5050\,\kms$,
with a dispersion (standard deviation) of $\sigma = 1240\,\kms$.
This fit is shown in Figure~\ref{si1260}.
Having started from the point of view that the \ion{Si}{2} was likely
to be unshocked, we were surprised to see that,
according to the fit, it is shocked \ion{Si}{2} which causes most of the
absorption,
although an appreciable quantity of unshocked Si is also present,
at velocities extending upwards from $5600\,\kms$.
The slight tail of \ion{Si}{2} absorption to high velocities
$> 7070\,\kms$ is naturally produced by the tail of the Gaussian profile
of the shocked \ion{Si}{2}.

The estimate (\ref{tion}) of the collisional ionization timescale of
\ion{Si}{2} presumed that all the \ion{Si}{2} 1260\,\AA\ absorption
was from unshocked \ion{Si}{2},
whereas the picture now is that only some of the absorption
is from unshocked \ion{Si}{2}.
According to the fit in Figure~\ref{si1260},
the optical depth of unshocked \ion{Si}{2} at the reverse shock front
is $\tau = 0.56 \pm 0.07$,
a little over half that adopted in estimate (\ref{tion}),
so a revised estimate of the ionization timescale of \ion{Si}{2} is
not quite double that of the original estimate (\ref{tion}):
\be
\label{tion'}
	{t_\SiII \over t} =
	2.5 {n_\SiII \over n_e}
	\ .
\ee
Evidently the conclusion remains that the ionization timescale of \ion{Si}{2}
is comparable to the age of the remnant.

\subsection{Shock jump conditions}
\label{jump}

The fitted profile of the \ion{Si}{2} 1260\,\AA\ feature in Figure~\ref{si1260}
includes both unshocked and shocked components.
The consistency of the fitted parameters can be checked against
the jump conditions for a strong shock.
The shock jump conditions predict that
the three-dimensional velocity dispersion $3^{1/2} \sigma$ of the ions
should be related to
the deceleration $\Delta v$ of the shocked gas
by energy conservation
\be
\label{Dv}
	3^{1/2} \sigma
	=
	\Delta v
\ee
provided that all the shock energy goes into ions.
The observed dispersion is
\be
\label{sigmaobs}
	3^{1/2} \sigma = 3^{1/2} \times ( 1240 \pm 40 \,\kms )
	= 2140 \pm 70 \,\kms
\ee
while the observed deceleration is
\ba
\label{Dvobs}
	\Delta v &=& ( 7070 \pm 50 \,\kms ) - ( 5050 \pm 60 \,\kms )
	\nn
	&=& 2020 \pm 80 \,\kms
	\ .
\ea
These agree remarkably well,
encouraging us to believe that this interpretation is along the right lines.
The reverse shock velocity, $v_s$, corresponding to the observed dispersion is
\be
\label{vs}
	v_s = (16/3)^{1/2} \sigma = 2860 \pm 100 \,\kms
	\ .
\ee
We prefer to infer the shock velocity from the observed dispersion rather than
from the observed deceleration $\Delta v$,
since the latter may underestimate the true deceleration, if, as is likely,
the shocked Si is moving on average slightly faster than the immediate
postshock gas (see below).

The predicted equality (\ref{Dv}) between the deceleration and ion dispersion
holds provided
that all the shock energy goes into ions,
and that the bulk velocity and dispersion of the ions in the shocked gas
are equal to their postshock values.
Since the shocked \ion{Si}{2} can last for a time comparable to the age
of the remnant (equation [\ref{tion'}]),
it cannot be assumed automatically that the bulk velocity and dispersion of the
observed \ion{Si}{2} ions are necessarily equal to those immediately
behind the reverse shock front.
We discuss first the issue of the bulk velocity,
then the dispersion,
and finally the question of collisionless heating
in subsection \ref{collisionless}.

Consider first the bulk velocity of the shocked ions.
In realistic hydrodynamic models,
the velocity of shocked gas increases outward from the reverse shock.
Indeed, the fact that
the observed dispersion $3^{1/2} \sigma$
is larger than
the observed deceleration $\Delta v$
by $120 \pm 130 \,\kms$
(the uncertainty here takes into account the correlation between
the uncertainties in $\sigma$ and $\Delta v$)
is consistent with the notion that the shocked \ion{Si}{2} is moving on average
$120 \pm 130 \,\kms$ faster than the immediate postshock gas.
This modest velocity is consistent with expectations from
one-dimensional hydrodynamic simulations appropriate to SN1006
(see HF88, Fig.~2),
according to which
shocked ejecta remain relatively close to the reverse shock front.
However,
the deceleration of ejecta is generally Rayleigh-Taylor unstable
(e.g.\ Chevalier, Blondin \& Emmering 1992),
which instabilities could have caused shocked \ion{Si}{2} to appear
at velocities many hundred $\kms$ faster than the immediate postshock gas.
Since the observations do not show this,
it suggests, though by no means proves,
either that Rayleigh-Taylor instabilities are not very important,
or perhaps that the line of sight through SN1006 to the SM star
happens to lie between Rayleigh-Taylor plumes.

What about the ion dispersion?
If there were a range of ion dispersions in the shocked gas,
then the line profile would tend to be more peaked and have broader wings
than a simple Gaussian.
The observed profile of the blue edge of the redshifted \ion{Si}{2} 1260\,\AA\ 
feature is consistent with a Gaussian, which suggests, again weakly,
that the ion dispersion does not vary by a large factor
over the bulk of the shocked \ion{Si}{2}.

While the observations agree well with the simplest possible interpretation,
it is certainly possible to arrange situations
in which a combination of Rayleigh-Taylor instabilities,
spatially varying ion dispersion,
and collisionless heating conspire to produce fortuitous agreement
of the observations with the jump condition (\ref{Dv}).

\subsection{Collisionless heating}
\label{collisionless}

The shock jump condition (\ref{Dv}) is valid provided that all the shock
energy goes into the ions,
rather than into electrons, magnetic fields, or relativistic particles.
The timescale for equilibration by Coulomb collisions between
electrons and ions or between ions and ions
is much longer than the age of SN1006.
However,
collisionless processes in the shock may also transfer energy,
and the extent to which these may accomplish equilibration remains uncertain
(see Laming et al.\ 1996 for a recent review).
The prevailing weight of evidence favors little if any collisionless heating
of electrons in the fast shocks in SN1006.
Raymond et al.\ (1995)
find similar velocity widths in emission lines of
\ion{H}{1} (Ly~$\beta$), \ion{He}{2}, \ion{C}{4}, \ion{N}{5}, and \ion{O}{6}
observed by HUT from the interstellar shock along the NW sector of SN1006.
They conclude that there has been little equilibration between ions,
though this does not rule out substantial electron heating.
From the same data,
Laming et al.\ (1996) 
argue that the ratio of \ion{C}{4} (which is excited mainly by protons)
to \ion{He}{2} (which is excited mainly by electrons)
is again consistent with little or no electron-ion equilibration.
Koyama et al.\ (1995) present spectral and imaging evidence from ASCA
that the high energy component of the x-ray spectrum of SN1006
is nonthermal synchrotron radiation,
obviating the need for collisionless electron heating
(see also Reynolds 1996).

The results reported here, equations (\ref{sigmaobs}) and (\ref{Dvobs}),
tend to support the conclusion
that virtually all the shock energy is deposited into the ions.
If some of the shock energy were chaneled into electrons,
magnetic fields, or relativistic particles,
then the ion dispersion would be lower than predicted by
the observed deceleration,
whereas the opposite is observed --- the ion dispersion is slightly higher.

\subsection{Column density and mass of \protect\ion{Si}{2}}
\label{mass}

The numbers given here are summarized in Table~\ref{redtab}.
Quoted uncertainties in column densities and masses here and
throughout this paper ignore uncertainties in oscillator strengths.
The uncertainties in the oscillator strengths of \ion{Si}{2} 1260\,\AA\
is 17\%, and of \ion{Si}{3} 1206\,\AA\
and \ion{Si}{4} 1394, 1403\,\AA\ are 10\% (Morton 1991).

The column density of shocked and unshocked \ion{Si}{2}
follows from integrating over the line profiles shown in
Figure~\ref{si1260}.
The column density $N_\SiII^{\rmn sh}$ of shocked \ion{Si}{2} is
\be
\label{NSiIIshk}
	N_\SiII^{\rmn sh}
	= {m_e c \over \pi e^2 f \lambda} (2\pi)^{1/2} \sigma \tau_0
	= 9.0 \pm 0.3 \times 10^{14} \cm^{-2}
\ee
where $\tau_0 = 0.98 \pm 0.02$ is the optical depth at line center of the
fitted Gaussian profile of the shocked \ion{Si}{2},
and the factor $(2\pi)^{1/2} \sigma$ comes from integrating over
the Gaussian profile.
The profile of unshocked \ion{Si}{2} is the residual after
the shocked \ion{Si}{2} is subtracted from the total,
and we measure its column density $N_\SiII^{\rmn unsh}$ by integrating
the unshocked profile over the velocity range $5600$-$7200 \,\kms$:
\be
\label{NSiIIunshk}
	N_\SiII^{\rmn unsh}
	= {m_e c \over \pi e^2 f \lambda} \!\int\! \tau^{\rmn unsh} \dd v
	= 1.5 \pm 0.2 \times 10^{14} \cm^{-2}
\ee
where the uncertainty is largely from uncertainty in the fit
to the shocked \ion{Si}{2}.
Integrating the full \ion{Si}{2} profile over $1000$-$8200 \,\kms$
yields the total column density of \ion{Si}{2} 
\be
\label{NSiIItot}
	N_\SiII
	= 10.5 \pm 0.1 \times 10^{14} \cm^{-2}
\ee
where the uncertainty is from photon counts,
and is smaller than the uncertainties in either of the shocked or
unshocked \ion{Si}{2} column densities individually
(because there is some uncertainty in allocating the total column density
between the shocked and unshocked components).

The corresponding masses of unshocked and shocked \ion{Si}{2}
can also be inferred,
if it is assumed that the Si was originally ejected spherically symmetrically.
In subsection \ref{blue}
we will argue that the absence of blueshifted \ion{Si}{2} absorption can be
explained if the reverse shock has passed entirely through the Si on the
near side of SN1006, and Si has been collisionally ionized to high ion stages.
Thus the absence of blueshifted \ion{Si}{2} need not conflict with
spherical symmetry of Si ejected in the supernova explosion.
If the shocked ejecta are taken to lie in a thin shell at a
free expansion radius of $v = 7200 \,\kms$
(slightly outside the position of reverse shock ---
cf.\ the argument in paragraph 3 of subsection \ref{jump}),
then the mass of shocked \ion{Si}{2} is
\be
\label{MSiIIshk}
	M_\SiII^{\rmn sh}
	= 4\pi m_\SiII (v t)^2 N_\SiII^{\rmn sh}
	= 0.127 \pm 0.006 \,\Msun
\ee
where the uncertainty includes only the uncertainty in the column density
of shocked \ion{Si}{2}.
This value should perhaps be modified to
$M_\SiII^{\rmn sh} = 0.13 \pm 0.01 \,\Msun$
to allow for uncertainty in the radial position of the shocked \ion{Si}{2}.
The mass of unshocked \ion{Si}{2} is an integral over the unshocked
line profile
\be
	M_\SiII^{\rmn unsh}
	= 4\pi m_\SiII t^2 \!\int\! v^2 \dd N_\SiII^{\rmn unsh}
	= 0.017 \pm 0.002 \,\Msun
\ee
where the uncertainty is largely from uncertainty in the fit
to the shocked \ion{Si}{2}.

\subsection{Purity of shocked Si}
\label{purity}

We show below
that the observed column density of \ion{Si}{2}
is close to the column density which would be predicted under the simple
assumption of steady state collisional ionization downstream of the shock
(note that recombination is negligible).
Now the assumption of steady state ionization is surely false,
since the timescale to ionize \ion{Si}{2} is comparable to the age of
the remnant, equation (\ref{tion'}).
Below we will argue that the effect of non-steady state is generally such
as to reduce the column density below the steady state value.
The constraint that the observed column density should be less than
or comparable to the steady state value then leads to an upper limit on the
electron to ion ratio, equation (\ref{nela}).
The fact that this upper limit is not much greater than unity
leads to the interesting conclusion that the bulk of the observed
shocked \ion{Si}{2} must be fairly pure,
since admixtures of other elements would increase the electron to ion ratio
above the limit.

If the postshock number density of \ion{Si}{2} ions is $n_\SiII$
(which is 4 times the preshock number density),
then at shock velocity $v_s$
the number of \ion{Si}{2} ions entering the reverse shock per unit area
and time is
$n_\SiII v_s/4$.
The \ion{Si}{2} ions are collisionally ionized by electrons in the shocked gas
at a rate $n_e \langle \sigma v \rangle_\SiII$ ionizations per unit time.
For $v_s = 2860 \,\kms$ and
$\langle \sigma v \rangle_\SiII = 6.1 \times 10^{-8} \cm^3 \s^{-1}$,
it follows that
the column density $N_\SiII^{\rmn steady}$ of shocked \ion{Si}{2}
in steady state should be
\be
\label{NSiIIsteady}
	N_\SiII^{\rmn steady} =
	{n_\SiII v_s \over 4 n_e \langle \sigma v \rangle_\SiII}
	= 1.2 \times 10^{15} \cm^{-2} {n_\SiII \over n_e}
	\ .
\ee
If, as will be argued below,
the actual (non-steady state) column density (\ref{NSiIIshk})
of shocked \ion{Si}{2}
is less than or comparable to the steady state value (\ref{NSiIIsteady}),
then the mean ratio of electron to \ion{Si}{2} density in the shocked gas
satisfies
\be
\label{nela}
	{n_e \over n_\SiII} \la 1.3
	\ .
\ee
The ratio must of course also satisfy $n_e / n_\SiII \ge 1$,
since each \ion{Si}{2} ion itself donates one electron.
Such a low value (\ref{nela}) of the electron to \ion{Si}{2} ratio in the
shocked \ion{Si}{2} would indicate that the bulk of the shocked \ion{Si}{2}
must be of a rather high degree of purity,
since the presence of significant quantities of other elements or
higher ionization states of Si would increase the number of electrons
per \ion{Si}{2} ion above the limit (\ref{nela}).
Indeed, even if the \ion{Si}{2} entering the shock were initially pure,
ionization to \ion{Si}{3} and higher states would release
additional electrons.
In steady state, the mean electron to \ion{Si}{2} ratio experienced
by \ion{Si}{2} during its ionization from an initially pure \ion{Si}{2}
state is $n_e / n_\SiII = 1.5$,
already slightly larger than the limit (\ref{nela}).

The limit (\ref{nela}) on the electron to \ion{Si}{2} ratio
is so low as to make it difficult to include even modest quantities
of other elements with the silicon.
This may be a problem.
While Si is generally the most abundant element
in Si-rich material produced by explosive nucleosynthesis,
other elements, notably sulfur, usually accompany the silicon.
For example, the deflagrated white dwarf model W7 of
Nomoto et al.\ (1984)
contains $0.16 \,\Msun$ of Si
mostly in a layer which is about 60\% Si, 30\% S by mass.
At this elemental abundance,
and assuming similar ionization states for all elements,
the expected mean electron to \ion{Si}{2} ratio in steady state would be
$n_e / n_\SiII = 1.5 / 0.6 = 2.5$,
almost twice the limit given by equation (\ref{nela}).
Because of this potential difficulty, we discuss carefully below
how robust is the constraint (\ref{nela}).
First we address the accuracy of the predicted value (\ref{NSiIIsteady})
of the steady state column density,
and then we discuss the non-steady state case.

The electron to \ion{Si}{2} ratio could be increased
if the predicted steady state column density (\ref{NSiIIsteady}) were increased,
either by increasing the shock velocity $v_s$,
or by reducing the collisional ionization rate
$\langle \sigma v \rangle_\SiII$.
We consider the former first, then the latter.
The shock velocity $v_s = 2860 \,\kms$ is inferred from the observed
ion dispersion in the \ion{Si}{2} 1260\,\AA\ line, equation (\ref{vs}).
This should give a fair estimate of the mean shock velocity of the
observed shocked \ion{Si}{2},
except for a small correction from the fact that
the ion dispersion (temperature) in the past would have been higher because
of adiabatic expansion of the remnant.
Moffet, Goss \& Reynolds (1993)
find that the global radius $R$ of the radio remnant is currently increasing
with time according to $R \propto t^{0.48 \pm 0.13}$.
If the ambient ISM is assumed to be uniform,
this indicates that the pressure in the remnant is varying with time as
$P \propto ( R / t )^2 \propto t^{-1.04 \pm 0.26}$,
hence that the temperature is varying in Lagrangian gas elements as
$T \propto P^{2/5} \propto t^{-0.42 \pm 0.10}$,
hence that the ion dispersion is varying as
$\sigma \propto T^{1/2} \propto t^{-0.21 \pm 0.05}$,
a rather weak function of time.
If one supposes that the observed \ion{Si}{2} was shocked
on average when SN1006 was say half its current age,
then the dispersion, hence the reverse shock velocity,
could have been $\sim 20\%$ higher than at present.
The steady state column density (\ref{NSiIIsteady}) would then be
$\sim 20\%$ higher,
and the constraint on the electron to \ion{Si}{2} ratio (\ref{nela})
would be relaxed slightly to
$n_e / n_\SiII \la 1.5$.

The collisional ionization rate
$\langle \sigma v \rangle_\SiII = 6.1 \times 10^{-8} \cm^3 \s^{-1}$
used in equation (\ref{NSiIIsteady}) comes from integrating the
cross sections of Lennon et al.\ (1988)
over a Maxwellian distribution of electrons at a temperature of 83\,eV.
The quoted error in the cross-sections is 60\%,
the large uncertainty arising from the fact
that the cross-sections for \ion{Si}{2} are derived from
isoelectronic scaling rather than from real data.
Reducing the ionization rate by 0.2 dex
would relax the constraint (\ref{nela}) to
$n_e / n_\SiII \la 2.0$.

The temperature of 83\,eV used in the collisional ionization rate above
is the temperature reached by electrons as a result
of Coulomb collisions with \ion{Si}{2} ions
over the collisional ionization timescale of \ion{Si}{2}.
The assumption here that there is no collisionless heating of electrons
in the shock is in accordance with the arguments given
in subsection \ref{collisionless}.
Actually,
the ionization rate of \ion{Si}{2} by electron impact,
as derived from the cross-sections of Lennon et al.,
has a broad maximum at an electron temperature of $\sim 200$\,eV,
and varies over only $5$-$7 \times 10^{-8} \cm^3 \s^{-1}$
for electron temperatures 40-1000\,eV.
Thus uncertainty in the electron temperature does not lead to much
uncertainty in the collisional ionization rate,
unless there is substantial collisionless heating
of electrons to temperatures much higher than 1\,keV.

We have argued against significant collisionless electron heating
in subsection~\ref{collisionless}.
However, if
collisionless electron heating to temperatures greater than 1\,keV did occur,
it would imply both a higher shock velocity,
since the observed ion dispersion would underestimate the shock energy,
and a lower collisional ionization rate,
both of which act to increase the steady state column density
(\ref{NSiIIsteady}).
Thus collisionless electron heating, if it occurs,
would allow a larger electron to \ion{Si}{2} ratio than
given by equation (\ref{nela}).

We now turn to the argument that in a non-steady state situation, as here,
the column density of shocked \ion{Si}{2} is likely to be less
than the steady state column density (\ref{NSiIIsteady}),
which leads to the constraint (\ref{nela}) on the
mean electron to \ion{Si}{2} ratio in the shocked \ion{Si}{2}.
In the first place,
simply truncating the Si at some point downstream of the shock
will give a lower column density than the steady state value.
Secondly,
geometric effects tend to reduce the column density below the steady
state value.
That is,
the column density of \ion{Si}{2} is diluted by the squared ratio
$(r_s/r)^2$ of the original radius $r_s$ of the gas at the time it was
shocked to the present radius $r$ ($> r_s$) of this shocked gas.
Thirdly,
if the density profile of shocked Si at the present time increases outwards,
then the faster ionization of the denser, earlier shocked, gas
reduces its column density per interval of ionization time,
and the net column density is again lower than steady state
(a more rigorous demonstration of this is given in the Appendix).
Conversely,
if the density profile of shocked Si decreases outwards,
then the net column density can be higher than steady state,
but only if the flow is allowed to continue for
sufficiently longer than a collisional ionization time.
However, according to equation (\ref{tion'})
the ionization timescale at the present \ion{Si}{2} density
is comparable to the age of the remnant,
and the ionization timescale would be longer at lower density,
so there is not much room for increasing the column density this way either.

So is there any way that the actual column density of \ion{Si}{2}
could be higher than the steady state value?
Clearly yes, given sufficient freedom with the density profile of shocked Si.
For example,
one possibility is that there is a `hump' in the shocked
\ion{Si}{2} density profile, such that the density increases outward of the
present position of the reverse shock front, but then declines at larger radii.
Some tuning is required to ensure that
the density on both near and far sides of the hump
is high enough to produce significant column density,
but not so high as to ionize Si above \ion{Si}{2}.
The higher the column density, the more fine-tuning is required.
%Any such hump in the shocked \ion{Si}{2} should not be too dramatic,
%else the ion temperature would vary
%(the pressure being constant),
%and the resulting superposition of ion dispersions in the shocked \ion{Si}{2}
%would tend
%(as argued already in the penultimate paragraph of subsection~\ref{jump})
%to produce a peakier line profile than the observed Gaussian profile
%of shocked \ion{Si}{2} in the 1260\,\AA\ feature,
%Figure~\ref{si1260}.

We thus conclude that while some violation of the limit
(\ref{nela}) on the mean electron to \ion{Si}{2} ratio in the shocked
\ion{Si}{2} is possible,
greater violations, exceeding say a factor of two, are less likely.
It then follows that the bulk of the shocked \ion{Si}{2}
is likely to be of a fairly high degree of purity.
In particular,
there is unlikely to be much iron mixed in with the shocked silicon,
a conclusion which is consistent with the absence of \ion{Fe}{2}
absorption with the same profile as the shocked \ion{Si}{2},
as discussed in subsection~\ref{shockedFe}.
To avoid misunderstanding,
this statement refers only to the shocked \ion{Si}{2}:
iron could be mixed with the unshocked Si,
and indeed the absorption profile of \ion{Fe}{2},
Figure~\ref{rho} below,
does suggest that there is some Fe mixed with unshocked Si.

\subsection{\protect\ion{Si}{3} and \protect\ion{Si}{4} line profiles}
\label{SiIII+IV}

Given that shocked \ion{Si}{2} apparently persists
for a time comparable to its ionization time,
it is difficult to avoid producing an appreciable quantity of
\ion{Si}{3} and \ion{Si}{4} as the result of collisional ionization of
\ion{Si}{2} in the shocked ejecta.
We thus conclude that it is likely that most of the observed
\ion{Si}{3} and \ion{Si}{4}
absorption arises from shocked ejecta.
This is consistent with the observed line profiles,
as will now be discussed.

\begin{figure}[tb]
\epsfbox[170 262 415 525]{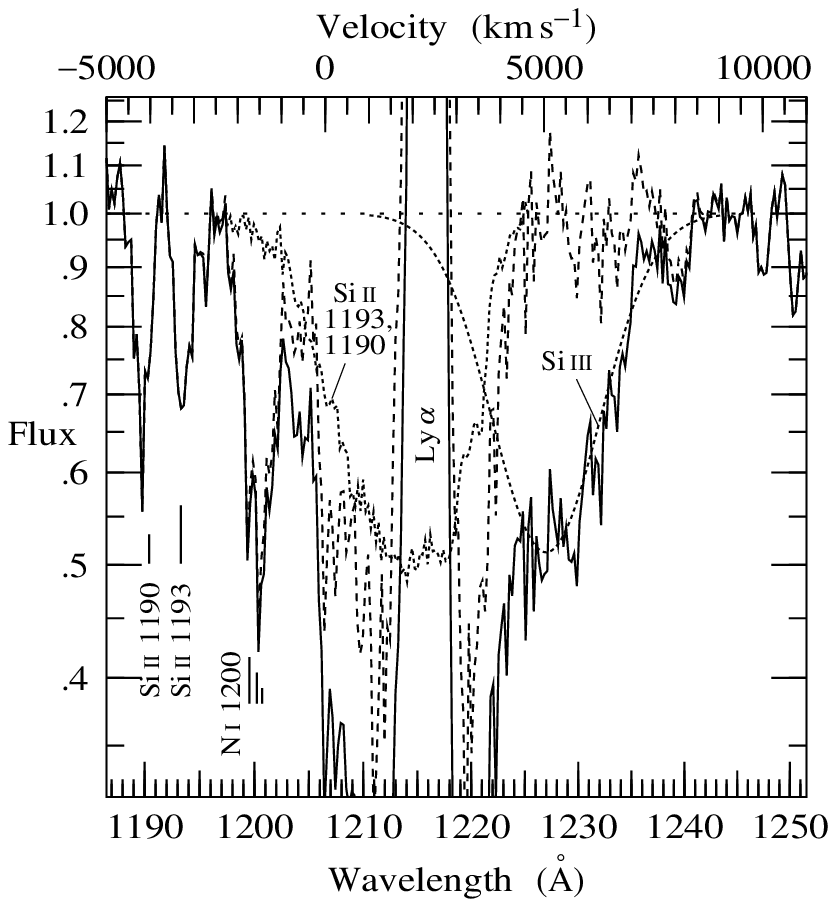}
%\vbox to94mm{\rule{0pt}{94mm}}
%\special{psfile=si3.ps angle=0 hoffset=-175 voffset=-265 vscale=100 hscale=100}
  \caption[1]{
HST G130H spectrum (solid line) relative to the adopted stellar continuum
around the redshifted \ion{Si}{3} 1206.500\,\AA\ 
($f = 1.669$) feature.
Upper axis shows velocity in the rest frame of the \ion{Si}{3} line.
The position and width of the fitted Gaussian profile of \ion{Si}{3}
(dotted line)
have been constrained to be the same as that of the
\ion{Si}{2} 1260\,\AA\ feature.
Also shown is the residual (dashed line) after subtraction both of
the \ion{Si}{3} fitted Gaussian,
and of the contribution of redshifted
\ion{Si}{2} 1193, 1190\,\AA\ (dotted line)
assumed to have the same profile as the \ion{Si}{2} 1260\,\AA\ feature.
The residual shows,
besides Ly\,$\alpha$ emission and absorption,
a weak indication of absorption by unshocked \ion{Si}{3}
at velocities $5500$-$7000 \,\kms$.
The adopted stellar continuum is the same as that
for the \ion{Si}{2} 1260\,\AA\ feature in Figure~\ref{si1260}.
\label{si3}
}
\end{figure}

\begin{figure}[tb]
\epsfbox[170 262 415 525]{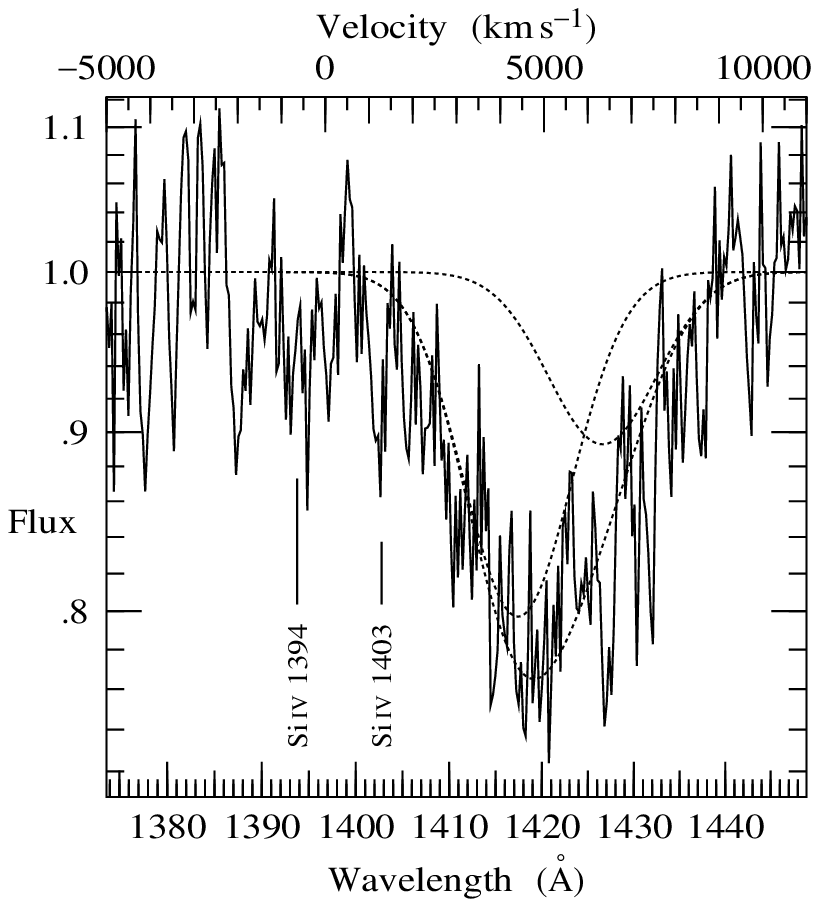}
%\vbox to94mm{\rule{0pt}{94mm}}
%\special{psfile=si4.ps angle=0 hoffset=-175 voffset=-265 vscale=100 hscale=100}
  \caption[1]{
HST G130H spectrum relative to the adopted stellar continuum around the
redshifted \ion{Si}{4} 1393.755, 1402.770\,\AA\ ($f = 0.5140$, 0.2553) feature.
Upper axis shows velocity in the rest frame
of the weighted mean wavelength of the feature.
Dotted lines show the best fit to a Gaussian pair with line center and
dispersion constrained to be that of \ion{Si}{2} 1260\,\AA\ for each
component of the doublet,
and with optical depths fixed equal to the ratio of oscillator strengths
times wavelengths of the doublet.
The best fit dispersion of the \ion{Si}{4} is $1700 \pm 100 \,\kms$,
which is $4.5 \sigma$ larger than the $1240 \,\kms$ dispersion
of the fit shown here.
There is no evidence of unshocked \ion{Si}{4}.
The adopted stellar continuum is
$\log F =
\log ( 2.9 \times 10^{-14} \,\erg\,\s^{-1}\,\cm^{-2}\,{\rmn \AA}^{-1} )
- 1.4 \log (\lambda/1397\,{\rmn \AA})$.
\label{si4}
}
\end{figure}

Figures~\ref{si3} and \ref{si4} show fits to
the redshifted
\ion{Si}{3} 1206\,\AA\ 
and \ion{Si}{4} 1394, 1403\,\AA\ features
using as templates the shocked and unshocked profiles of
\ion{Si}{2} 1260\,\AA\ shown in Figure~\ref{si1260}.
Table~\ref{sitab} gives
the fitted column densities of shocked and unshocked
\ion{Si}{3} and \ion{Si}{4},
expressed relative to the best fit column density of shocked and unshocked
\ion{Si}{2} given in Table~\ref{redtab}.

The \ion{Si}{3} 1206\,\AA\ profile appears to be mostly shocked.
There is some indication of unshocked \ion{Si}{3}
over $5500$-$7000 \,\kms$ at the $2 \sigma$ level,
Table~\ref{sitab},
as suggested by the residual profile after subtraction of shocked \ion{Si}{3}
plotted in Figure~\ref{si3}.
The dispersion of the fitted Gaussian profile of the \ion{Si}{3},
if allowed to be a free parameter,
is $1290 \pm 60 \,\kms$ if unshocked \ion{Si}{3} is excluded,
or $1210 \pm 60 \,\kms$ if unshocked \ion{Si}{3} is admitted,
which are in good agreement with the $1240 \pm 40 \,\kms$ dispersion
of the \ion{Si}{2}.

The profile of the
\ion{Si}{4} 1394, 1403\,\AA\ feature, Figure~\ref{si4},
is consistent with containing no unshocked \ion{Si}{4}.
If the line center and width of the Gaussian pair fitted to
the \ion{Si}{4} 1394, 1403\,\AA\ doublet are allowed to be free,
then the center remains close to $5050 \,\kms$,
the same as for \ion{Si}{2} 1260\,\AA,
but the best fit dispersion of the \ion{Si}{4} is
$1700 \pm 100 \,\kms$,
which is $4.5 \sigma$ higher than the $1240 \,\kms$ dispersion of
\ion{Si}{2} 1260\,\AA.
The broader dispersion suggests that the \ion{Si}{4}
may be in slightly lower density gas than the \ion{Si}{2},
since the pressure is presumably the same for both.
It is not clear however that the observed difference in velocity width
between \ion{Si}{4} and \ion{Si}{2} is real.
One problem is that the continuum around \ion{Si}{4} appears less
well defined than for the \ion{Si}{2} and \ion{Si}{3} features.
As elsewhere in this paper,
we assume a stellar continuum which is linear in $\log F$-$\log \lambda$,
but in fact there is a hint of curvature, a large scale depression in the
continuum around \ion{Si}{4} 1394, 1403\,\AA\ (see WCHFLS96, Figure~1). 
If we have systematically misjudged the continuum, then it is possible that we
have underestimated the uncertainties in the parameters of \ion{Si}{4} given
Table~\ref{sitab}, perhaps by as much as a factor of 2.
The Gaussian profile shown in Figure~\ref{si4} is constrained to
have the same center and $1240 \,\kms$ dispersion as the
\ion{Si}{2} 1260\,\AA\ feature.
Visually, at least, the fit appears satisfactory.

%--------------------
% ApJ version
\begin{deluxetable}{lccc}
\tablewidth{0pt}
\tablecaption{Column densities of shocked and unshocked \protect\ion{Si}{3}
and \protect\ion{Si}{4}, relative to best fit column densities of
\protect\ion{Si}{2}
\label{sitab}}
\tablehead{& \colhead{\protect\ion{Si}{2}} & \colhead{\protect\ion{Si}{3}} &
\colhead{\protect\ion{Si}{4}}}
\startdata
Shocked & $1 \pm 0.03$ & $0.43 \pm 0.02$ & $0.41 \pm 0.02$ \nl
Unshocked & $1 \pm 0.13$ & $0.065 \pm 0.035$ & $0.02 \pm 0.07$ \nl
\enddata
\tablecomments{
Absolute column densities of \protect\ion{Si}{2} are given in
Table~\protect\ref{redtab}.
Column densities of \protect\ion{Si}{4} relative to \protect\ion{Si}{2}
are for fits in which the dispersion of \protect\ion{Si}{4} is constrained to
be that of \protect\ion{Si}{2} 1260\,\AA, namely $1240 \,\kms$.
The column densities of \protect\ion{Si}{4} become
$0.46 \pm 0.02$ (shocked) and $-0.06 \pm 0.06$ (unshocked)
if instead the dispersion of the \protect\ion{Si}{4} is taken to have the best
fit value $1700 \pm 100 \,\kms$.
}
\end{deluxetable}
%--------------------

%%--------------------
%% MNRAS version
%\begin{table}
%\caption{Column densities of shocked and unshocked \protect\ion{Si}{3}
%and \protect\ion{Si}{4}, relative to best fit column densities of
%\protect\ion{Si}{2}
%\label{sitab}}
%\begin{tabular}{@{}lccc}
%& \protect\ion{Si}{2} & \protect\ion{Si}{3} & \protect\ion{Si}{4} \\ [4pt]
%Shocked & $1 \pm 0.03$ & $0.43 \pm 0.02$ & $0.41 \pm 0.02$ \\
%Unshocked & $1 \pm 0.13$ & $0.065 \pm 0.035$ & $0.02 \pm 0.07$ \\ [5pt]
%\end{tabular}
%
%Absolute column densities of \protect\ion{Si}{2} are given in
%Table~\ref{redtab}.
%Column densities of \protect\ion{Si}{4} relative to \protect\ion{Si}{2}
%are for fits in which the dispersion of \protect\ion{Si}{4} is constrained to
%be that of \protect\ion{Si}{2} 1260\,\AA, namely $1240 \,\kms$.
%The column densities of \protect\ion{Si}{4} become
%$0.46 \pm 0.02$ (shocked) and $-0.06 \pm 0.06$ (unshocked)
%if instead the dispersion of the \protect\ion{Si}{4} is taken to have the best
%fit value $1700 \pm 100 \,\kms$.
%\end{table}
%%--------------------

\subsection{Total Si mass}
\label{Simass}

In subsection~\ref{purity} we showed that the observed column density
of shocked \ion{Si}{2} is close to the theoretical steady state value.
Is the same also true for \ion{Si}{3} and \ion{Si}{4}?
The answer is no.
In steady state,
the predicted column densities of shocked ionic species are inversely
proportional to their respective collisional ionization rates
$\langle \sigma v \rangle$,
modified by an appropriate electron to ion ratio
(cf.\ equation [\ref{NSiIIsteady}]).
Adopting rates
$6.1 \times 10^{-8} \cm^3 \s^{-1}$,
$2.2 \times 10^{-8} \cm^3 \s^{-1}$,
$1.1 \times 10^{-8} \cm^3 \s^{-1}$
for \ion{Si}{2}, \ion{Si}{3}, \ion{Si}{4} respectively
(Lennon et al.\ 1988),
and assuming a nominal $i$ electrons per ion for Si$^{+i}$,
yields relative column densities in steady state
\be
\label{NSisteady}
	N_\SiII : N_\SiIII : N_\SiIV
	= 1 : 1.4 : 1.8
	\ .
\ee
By comparison the observed shocked column densities are
$N_\SiII : N_\SiIII : N_\SiIV = 1 : 0.43 : 0.41$,
according to Table~\ref{sitab}.
Evidently
the observed abundances of shocked \ion{Si}{3} and \ion{Si}{4}
relative to \ion{Si}{2} are several times
less than predicted in steady state.

As discussed in subsection~\ref{purity},
there are several ways to reduce the column density below the steady state
value, of which the most obvious is to truncate the column density,
as is strongly suggested by the fact that the ionization timescales
of \ion{Si}{3} and \ion{Si}{4} are becoming long compared to the age
of the remnant.
In fact these ionization timescales are in precisely the same ratio
(cf.\ equation [\ref{tion}])
as the steady state column densities (\ref{NSisteady})
\be
\label{tionSi}
	t_\SiII : t_\SiIII : t_\SiIV
	= 1 : 1.4 : 1.8
\ee
and it has already been seen that the ionization timescale $t_\SiII$
of \ion{Si}{2} is comparable to the age of the remnant, equation (\ref{tion'}).

If it is true that it is the long ionization times which cause
the column densities of \ion{Si}{3} and \ion{Si}{4}
to be lower than steady state,
then this suggests that there may be little Si in higher ionization states
in the shocked gas on the far side of SN1006.
Thus it appears plausible that we are observing in UV absorption
essentially all the Si there is on the far side of SN1006
along the line of sight to the background SM star.
To avoid confusion, it should be understood that this statement refers
specifically to Si on the far side along this particular
line of sight.
Higher ionization states of Si are indicated by the observed
Si x-ray line emission
(Koyama et al.\ 1995),
which could arise from denser shocked gas in other parts of the remnant.

The mass of Si can be inferred from the observed column densities of
\ion{Si}{2}, \ion{Si}{3}, and \ion{Si}{4},
if it is assumed that silicon was ejected spherically symmetrically
by the supernova explosion.
We will argue in subsection~\ref{blue}
that the absence of blueshifted absorbing Si is not inconsistent
with spherical symmetry.
Taking the shocked and unshocked masses of \ion{Si}{2} from Table~\ref{redtab},
and the ratios of
\ion{Si}{2}, \ion{Si}{3}, and \ion{Si}{4}
from Table~\ref{sitab},
yields a total inferred Si mass of
\ba
	M_\Si
	&=& 0.127 ( 1 + 0.43 + 0.41 ) \Msun + 0.017 ( 1 + 0.065 ) \Msun
	\nn
	&=& 0.25 \pm 0.01 \,\Msun
	\ .
\label{MSi}
\ea
This is comparable to the $0.16\,\Msun$ of Si in model W7 of
Nomoto et al.\ (1984).
It is also consistent with the $0.20\,\Msun$ of Si inferred from
the strength of the Si\,K line observed with ASCA
(Koyama et al.\ 1995).
Koyama et al.\ do not quote a mass,
but they do state that the measured surface brightness of the Si\,K line
is 5 times higher than that in the model\footnote{
Hamilton et al.\ took the strength of the Si\,K line
to be (the upper limit to) that measured from
the Einstein Solid State Spectrometer
(Becker et al.\ 1980),
so there is a discrepancy between the ASCA and SSS data.
However, Hamilton et al.\ also noted in their Figure~6 a marked discrepancy
between the SSS and HEAO-1 data of Galas, Venkatesan \& Garmire (1982),
so it is reasonable to suspect an error in the normalization of the SSS data.
}
of Hamilton, Sarazin \& Szymkowiak (1986; see also Hamilton et al.\ 1985),
which had $0.04\,\Msun$ of Si.

\subsection{No blueshifted Si --- evidence for an inhomogeneous ISM?}
\label{blue}

There is no evidence for blueshifted Si absorption in the UV spectrum.
At best, there is a possible hint of a broad shallow depression
around $\sim - 5000 \,\kms$, from 1365\,\AA\ to 1390\,\AA,
on the blue side of the
\ion{Si}{4} 1394, 1403\,\AA\ line
(see WCHFLS96, Figure~1).
The possibility that some high velocity blueshifted \ion{Si}{2} 1260\,\AA\ 
is hidden in the red wing of \ion{Si}{3} 1206\,\AA\ is excluded by the
absence of corresponding blueshifted \ion{Si}{2} 1527\,\AA.

There are two possible reasons for the asymmetry in the observed Si absorption.
One is that there was an intrinsic asymmetry in the supernova explosion.
According to Garcia-Senz \& Woosley (1995),
the nuclear runaway that culminates in the explosion of a nearly
Chandrasekhar mass white dwarf begins as stable convective carbon burning,
and ignition is likely to occur off-center at one or more points
over a volume of the order of a pressure scale height.
The subsequent propagation of the convectively driven burning front is
Rayleigh-Taylor unstable
(Livne 1993;
Khokhlov 1995;
Niemeyer \& Hillebrandt 1995),
although Arnett \& Livne (1994) find that in delayed detonation models
the second, detonation, phase of the explosion tends to restore
spherical symmetry.

Thus asymmetry in the explosion is possible,
perhaps even likely, as an explanation of the asymmetry in the Si absorption,
especially since the absorption samples only a narrow line of sight
through SN1006 to the background SM star.
However, we do not pursue this possibility further,
in part because in abandoning spherical symmetry we lose any
predictive power,
and in part because there is another explanation which
is more attractive because it resolves some other observational problems.

The other possible cause of the asymmetry in the Si absorption
is that the ISM around SN1006 is inhomogeneous,
with the ISM on the far side of SN1006 having a significantly lower density
than the near side.
According to this hypothesis,
the low density on the far side is such that
the reverse shock on the far side has reached inward only to
a free expansion radius of $7070 \,\kms$,
whereas the higher density on the near side is such that
the reverse shock on the near side has passed entirely through the Si layer,
and Si has been collisionally ionized to stages higher than \ion{Si}{4},
making it unobservable in UV absorption.

A serious objection to the reverse shock on the near side being
farther in than the reverse shock on the far side
is the observation by WCFHS93
of blueshifted \ion{Fe}{2}
absorption certainly to velocities $-7000 \,\kms$,
perhaps to velocities $-9000 \,\kms$.
We will review the observational evidence for such high velocity
blueshifted \ion{Fe}{2} in Section \ref{iron} below,
where we will argue that the evidence is not compelling.

An inhomogeneous ISM around SN1006 is indicated by other observations.
Wu et al.\ (1983) and subsequent authors have remarked on the difficulty of
fitting the observed high velocity ($\sim 7000 \,\kms$) Si
within the confines of the observed interstellar blast wave,
if spherical symmetry is assumed.
At a distance of $1.8 \pm 0.3 \,\kpc$ (Laming et al.\ 1996),
the remnant's observed $15'$ radius
(Reynolds \& Gilmore 1986, 1993; see also Willingale et al.\ 1996)
at 980 years old\footnote{
Note there is a 23 year light travel time across one radius of the remnant,
so really we are seeing the far side at an age 23 years younger,
and the near side 23 years older, than the mean age.
}
corresponds to a free expansion radius
of $7700 \pm 1300 \,\kms$.
The difficulty is resolved if
the remnant of SN1006 bulges out on the far side,
because of the lower density there.

A second piece of evidence suggesting inhomogeneity
is the high $5050 \,\kms$ velocity of the shocked ejecta
behind the reverse shock on the far side of SN1006
inferred from the present observations,
compared with the $1800$-$2400 \,\kms$
(the lower velocity corresponds to no collisionless electron heating,
which is the preferred case)
velocity of shocked gas behind the interstellar shock
inferred from H\,$\alpha$ and UV emission line widths along the
NW sector of the remnant
(Kirshner et al.\ 1987;
Long et al.\ 1988;
Smith et al.\ 1991;
Raymond et al.\ 1995).
These two velocities, $5050 \,\kms$ versus $1800$-$2400 \,\kms$,
appear incompatible,
especially as the velocity of shocked gas is expected in realistic
hydrodynamic models to increase outwards from the reverse shock
to the interstellar shock
(see for example HF88, Fig.~2).
The incompatibility is resolved if the ISM around SN1006 is inhomogeneous,
with the density on the far side of SN1006 being substantially lower
than the density at the NW edge.

A final piece of evidence supporting inhomogeneity
comes from the observation of Si\,K line emission in x-rays
(Koyama et al.\ 1995).
This emission is likely to be from ejecta, since the inferred abundance
of Si is stated to be an order of magnitude higher than that of O, Ne, or Fe.
To ionize Si to high ionization states in the age of SN1006,
and to produce Si\,K line emission at the observed luminosity,
requires densities
$n_\Si \ga 10^{-2} \,\cm^{-3}$
(i.e.\ $n_e \ga 10^{-1} \,\cm^{-3}$ since the Si is highly ionized),
substantially higher than the postshock density of
$n_\SiII = 2 \times 10^{-4} \,\cm^{-3}$
deduced here from the redshifted \ion{Si}{2} 1260\,\AA\ absorption profile.
It is difficult to see how the required high Si density could be achieved
if the reverse shock everywhere around the remnant is at a radius as
high as $7070 \,\kms$,
whereas higher densities would occur naturally if
the reverse shock around most of the remnant were farther in,
since then ejecta would have been shocked at earlier times when
its density ($\propto t^{-3}$ in free expansion) was higher.

All these arguments point to the notion that the ISM density on the far
side of SN1006 is anomalously low compared to the density around the
rest of the remnant.

\subsection{Ionization of Si on the near side}
\label{blueion}

In Section~\ref{iron}, we will argue that the \ion{Fe}{2}
absorption profiles suggest that the reverse shock on the near side
of SN1006 may be at a free expansion radius of $4200 \,\kms$
(as with the Hubble expansion of the Universe, it is often convenient
to think in a comoving frame expanding freely with the ejecta,
so that a free expansion velocity $v = r/t$ can be thought of as a radius).
Here we estimate whether Si could have ionized beyond \ion{Si}{4},
as required by the absence of observed blueshifted Si absorption,
if the reverse shock on the near side is indeed at $4200 \,\kms$.

In our first estimate,
we find that blueshifted \ion{Si}{4} {\em should}\/ be observable,
with a column density $\sim 40\%$ that of the observed redshifted \ion{Si}{4}
absorption.
However, the column density is somewhat sensitive to the assumptions made,
and it is not hard to bring the column density down below observable
levels.

The line profile of any blueshifted absorbing Si would have
a width comparable to that of the observed broad redshifted Si absorption,
but the centroid would be shifted to lower velocities,
to $\sim -2000 \,\kms$,
if it assumed that the reverse shock velocity on the near side
is comparable to that, $v_s = 2860 \,\kms$, observed on the far side.
To avoid being observed at say $3 \sigma$,
the column density of blueshifted \ion{Si}{4} should be less than
$0.6 \times 10^{14} \,\cm^{-2}$.

To estimate the ionization of Si,
it is necessary to adopt a hydrodynamic model.
Now one interesting aspect of hydrodynamical models of deflagrated white dwarfs
expanding into a uniform ambient medium
is that the reverse shock velocity $v_s$ remains almost constant in time
(this conclusion is based on hydrodynamic simulations carried out by HF88).
In model W7 of Nomoto et al.\ (1984),
the reverse shock velocity varies (non-monotonically)
between $3300 \,\kms$ and $5200 \,\kms$
as it propagates inward from a free expansion radius of $13000 \,\kms$
to $700 \,\kms$,
after which the shock velocity accelerates.
Similarly in model CDTG7 (Woosley 1987, private communication),
which is similar to model CDTG5 of Woosley \& Weaver (1987),
the reverse shock velocity varies between $3200 \,\kms$ and $4100 \,\kms$
as it propagates inward from a free expansion radius of $10000 \,\kms$
to $1500 \,\kms$.
These numbers do not depend on the density of the ambient medium,
although they do depend on the ambient density being uniform.
If the reverse shock velocity $v_s$ remains constant in time,
then the radius $r$ of the reverse shock evolves with time $t$
according to $\dd r/\dd t = r/t - v_s$,
from which it follows that the free expansion radius $r/t$
of the reverse shock varies with time as
\be
\label{rt}
	{r \over t} = v_s \ln left ( {t_\ast \over t} right )
\ee
where $t_\ast$ is the age at which the reverse shock eventually hits
the center of the remnant.

The assumption that the reverse shock $v_s$ is constant in time
may not be correct for SN1006,
but it provides a convenient simplification to estimate the ionization of Si
on the near side.
Let us first estimate the ionization state of Si which was originally
at a free expansion radius of $7070 \,\kms$,
the current location of the reverse shock on the far side.
If the reverse shock on the near side is currently at a free expansion radius of
$r/t = 4200 \,\kms$,
and if it has moved at a constant $v_s = 2860 \,\kms$,
the measured value on the far side
(recall that the reverse shock velocity is independent of the ambient density,
for uniform ambient density),
then it would have passed through a free expansion radius of $7070 \,\kms$
when the age $t_s$ of SN1006 was
$t_s / t = \exp [ ( 4200 \,\kms - 7070 \,\kms ) / 2860 \,\kms ] = 0.37$
times its present age $t$,
according to equation (\ref{rt}).
The postshock density of Si ions at that time would have been
$(t / t_s)^3 = 20$ times higher than the presently observed postshock
density of $n_\SiII = 2.2 \times 10^{-4} \,\cm^{-3}$.
The ion density in the parcel of gas shocked at that time has been decreasing
because of adiabatic expansion.
The rate of decrease of density by adiabatic expansion can be inferred from the
observed global rate of expansion of the remnant
(Moffet, Goss \& Reynolds 1993)
\be
	R \propto t^\alpha
	\ ,\ \ \ 
	\alpha = {0.48 \pm 0.13}
	\ ,
\ee
which for an assumed uniform ambient density would imply that the pressure is
decreasing as
$P \propto ( R / t )^2 \propto t^{2 \alpha - 2}$,
hence that the density in Lagrangian gas elements is varying as
$n \propto P^{3/5} \propto t^{6(\alpha - 1)/5}$.
The current ionization time
$\tau \equiv \int_{t_s}^t n \,\dd t$
of the parcel of gas
originally at free expansion radius $7070 \,\kms$
which was shocked at an age $t_s / t = 0.37$
is then
\ba
	\tau
	&=& {n_\SiII t \over (6 \alpha - 1)/5} \left( {t \over t_s} \right)^2
	\left[ \left( {t \over t_s} \right)^{(6 \alpha - 1)/5} - 1 \right]
	\nn
	&=& 6.0 \times 10^7 \,\cm^{-3}\,\s
\label{tauf}
\ea
where $n_\SiII t = 6.6 \times 10^6 \,\cm^{-3}\,\s$ is the present
postshock density of \ion{Si}{2} ions times age
at the radius $7070 \,\kms$.
If the Si is assumed unmixed with other elements and initially singly ionized,
then the ionization time (\ref{tauf}) yields ion fractions
5\% \ion{Si}{3}, 34\% \ion{Si}{4}, and the remaining 61\% in higher stages.

Since this ionization state is close to (just past) the peak
in the \ion{Si}{4} fraction, it should be a good approximation
to estimate the column density of \ion{Si}{4} by expanding locally about
the conditions at $7070 \,\kms$.
The expected column density is
the steady state column density,
multiplied by a geometric factor,
and further multiplied by a `density profile' factor
$[1 + \gamma \tau / (n_s t_s)]^{-1}$,
as shown in the Appendix,
equation (\ref{dNdtau}).
The steady column density is
$N_\SiIV^{\rmn steady}$
= $n_\SiIV v_s / ( 4 n_e \langle \sigma v \rangle_\SiIV )$
= $22 \times 10^{14} \,\cm^{-2}$
assuming a nominal $n_e/n_\SiIV = 3$.
The geometric factor is
$(7070 t_s / 4200 t)^2 = 0.38$,
which is the squared ratio of the radius of the gas at the time $t_s$
it was shocked to its radius at the present time $t$.
For the density profile factor,
equation (\ref{tauf}) gives
$\tau/(n_s t_s) = [(t/t_s)^{(6\alpha-1)/5} - 1]/[(6\alpha-1)/5] = 1.22$,
while the logarithmic slope $\gamma = - \partial \ln n / \partial \ln t_s |_t$
of the shocked density profile, equation (\ref{gamma}), is
$\gamma$ =
$3 + v_s \partial \ln n^{\rmn unsh} / \partial (r/t) + 6(\alpha - 1)/5$
= $3.7$,
the 3 coming from free expansion,
the $v_s \partial \ln n^{\rmn unsh} / \partial (r/t) = 1.37$
from the observed unshocked \ion{Si}{2} density profile
(see eq.~[\ref{nSiquad}]) at $7070 \,\kms$
along with equation (\ref{rt}),
and the $6(\alpha - 1)/5 = -0.62$ from the reduction in density caused
by adiabatic expansion.
The resulting density profile factor
is $[1 + \gamma \tau / (n_s t_s)]^{-1} = 0.18$.
The expected column density of blueshifted \ion{Si}{4} is then 
$N_\SiIV = 1.5 \times 10^{14} \,\cm^{-2}$,
which is 40\% of the observed column density of redshifted \ion{Si}{4},
and 2.5 times the minimum ($3 \sigma$) observable column density.

Thus under a reasonable set of simplifying assumptions,
there should be an observable column density of blueshifted \ion{Si}{4},
contrary to observation.
However,
the expected column density is sensitive to the assumed parameters.
For example, reducing the shock velocity on the near side by 20\%
to $2300 \,\kms$ reduces the column density by a factor 2.5 to
the observable limit
$N_\SiIV = 0.6 \times 10^{14} \,\cm^{-2}$.
Whether the shock velocity on the near side is less or more than that
on the far side depends on the unshocked density profile of ejecta
(generally, a shorter exponential scale length of unshocked density with
velocity yields lower shock velocities).
The expected column density is also sensitive to the shocked density profile,
as might be guessed from the fact that the density profile factor of $0.18$
estimated above differs substantially from unity.

Alternatively,
the column density of \ion{Si}{4} could be reduced below the
observable limit by mixing the Si with a comparable mass of other elements,
such as iron, since this would increase the number of electrons
per silicon ion, causing more rapid ionization.
The possibility that there is iron mixed with Si at velocities
$\la 7070 \,\kms$ gains some support from the observed density
profile of \ion{Fe}{2}, shown in Figure~\ref{rho} below.
Note this does not conflict with the argument in subsection~\ref{purity}
that most of the shocked Si
(which was originally at higher free expansion velocities)
is probably fairly pure,
with little admixture of other elements such as iron.

\section{Iron}
\label{iron}

We have argued above that an attractive explanation for
the presence of redshifted Si absorption and absence of blueshifted absorption
is that the ISM on the near side of SN1006 is much denser than that on the
far side,
so that a reverse shock has already passed all the way through the Si layer
on the near side, ionizing it to high ionization stages,
whereas the reverse shock is still moving through the Si layer on the far side.
This picture appears to conflict with our previously reported result
(WCFHS93),
according to which blueshifted \ion{Fe}{2} is present to velocities
$\sim - 8000 \,\kms$.
In this Section we reexamine the \ion{Fe}{2} absorption lines to see
how robust is this result.

In the average broad \ion{Fe}{2} profile shown in Figure~3 of WCFHS93,
redshifted \ion{Fe}{2} seems to extend up to about $7000 \,\kms$,
but not much farther.
This is consistent with the argument of the present paper,
which is that the reverse shock on the far side of SN1006 lies at
$7070 \,\kms$.
The problem lies on the blueshifted side of the \ion{Fe}{2} profile,
which appears to extend clearly to $- 7000 \,\kms$,
possibly to $- 9000 \,\kms$.

Figure~2 of WCFHS93 shows separately the two broad
\ion{Fe}{2} 2383, 2344, 2374\,\AA\ and
\ion{Fe}{2} 2600, 2587\,\AA\ features.
The \ion{Fe}{2} 2600, 2587\,\AA\ feature
appears to have a fairly abrupt blue edge,
although there is perhaps a tail to higher velocities
depending on where the continuum is placed.
The blue edge is at a velocity of $-4200 \,\kms$
with respect to the stronger 2600\,\AA\ component of the doublet,
and the same edge appears at this velocity in the average \ion{Fe}{2}
profile shown in WCFHS93's Figure~3.
We will argue that this edge plausibly represents the position of
the reverse shock on the near side of SN1006.

In contrast to \ion{Fe}{2} 2600\,\AA,
the deconvolved profile of the \ion{Fe}{2} 2383\,\AA\ feature,
plotted in the bottom curve of WCFHS93's Figure~2,
shows blueshifted absorption clearly extending to
$\la 7000 \,\kms$.
We note that the second strongest component of the triplet, 2344\,\AA,
with $1/3$ the oscillator strength of the principal 2383\,\AA\ component,
lies at $- 4900 \,\kms$ blueward of the principal line,
and uncertainty involved in removing the secondary component in the
deconvolution procedure could tend to obscure any sharp blue edge at
$- 4200 \,\kms$ on the principal component.

\subsection{\protect\ion{Fe}{2} analysis}
\label{reanalysis}

\begin{figure*}[tb]
\begin{minipage}{175mm}
\epsfbox[25 256 542 513]{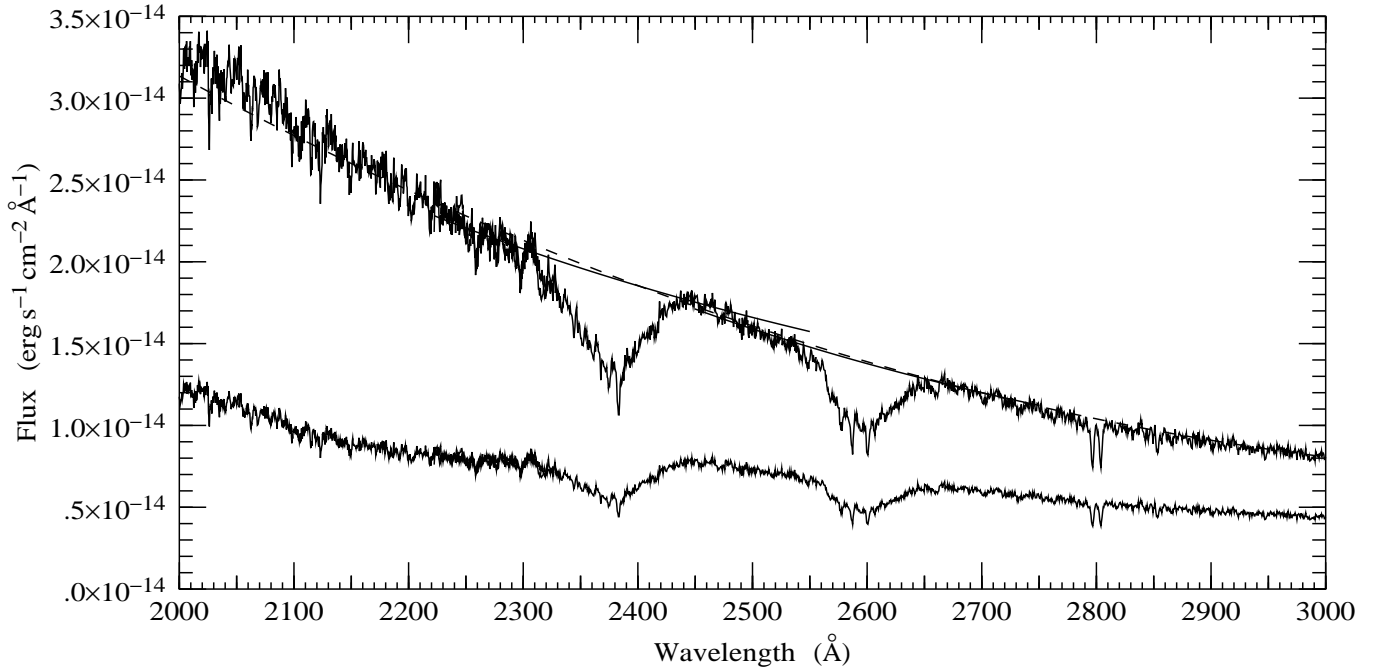}
%\vbox to98mm{\rule{0pt}{98mm}}
%\special{psfile=fluxfe.ps
%         angle=0 hoffset=-57 voffset=-255 vscale=100 hscale=100}
  \caption[1]{
G190H and G270H spectra,
with the calibrated spectra at bottom,
and the dereddened spectra at top.
Also shown are
the continua linear in $\log F$-$\log \lambda$ (solid lines)
adopted in the present paper
for each of the \ion{Fe}{2} 2600\,\AA\ and \ion{Fe}{2} 2383\,\AA\ features,
and a continuum quadratic in $F$-$\lambda$ (dashed line) similar
(but not identical, because of the slightly different reddening)
to that adopted by WCFHS93.
The linear continua are
$\log F =
\log ( 1.89 \times 10^{-14} \,\erg\,\s^{-1}\,\cm^{-2}\,{\rmn \AA}^{-1} )
- 2.7 \log (\lambda/2383\,{\rmn \AA})$
for the \ion{Fe}{2} 2383\,\AA\ feature, and
$\log F =
\log ( 1.38 \times 10^{-14} \,\erg\,\s^{-1}\,\cm^{-2}\,{\rmn \AA}^{-1} )
- 3.65 \log (\lambda/2600\,{\rmn \AA})$
for the \ion{Fe}{2} 2600\,\AA\ feature.
\label{fluxfe}
}
\end{minipage}
\end{figure*}

In this subsection we present details of a reanalysis of the
\ion{Fe}{2} absorption lines in the HST G190H and G270H spectra
originally analyzed by WCFHS93.
The observations are described by WCFHS93,
and here we describe the differences between
the analysis here and that of WCFHS93.
In carrying out the reanalysis,
we paid particular attention to procedures
which might affect the blue wing of the \ion{Fe}{2} 2383\,\AA\ feature.

The G190H and G270H spectra overlap over the wavelength range
2222-2330\,\AA,
and we merged the two spectra in this region using inverse variance weighting,
whereas WCFHS93 chose to abut the spectra at 2277\,\AA.
In merging the spectra we interpolated the G270H data to the same bin size
as the G190 spectrum, which has higher resolution
(2\,\AA\ versus 2.8\,\AA),
and higher signal to noise ratio than the G270H spectrum
in the overlap region.
According to the FOS handbook,
there is contamination at the level of a few percent in the G190H spectrum
above 2300\,\AA\ from second order,
but, given the absence of strong features over 1150-1165\,\AA,
we accept this contamination in the interests of obtaining higher signal
to noise ratio.
The merged spectrum is noticeably less choppy than the G270H spectrum alone
in the overlap region.

The 2200\,\AA\ extinction bump is close to the blue wing of
the \ion{Fe}{2} 2383\,\AA\ feature,
so we re-examined the reddening correction.
In practice, the changes made here
had little effect on the profile of the \ion{Fe}{2} 2383\,\AA\ feature.
We dereddened the G190H and G270H spectra using the extinction curve of
Cardelli, Clayton \& Mathis (1989),
adopting $E_{B-V} = 0.119$, which is the best fitting value
determined by Blair et al.\ (1996) from HUT data,
and $R \equiv A_V / E_{B-V} = 3.0$.
The value of $E_{B-V}$ is slightly higher than the value
$E_{B-V} = 0.1 \pm 0.02$ measured by WCFHS93
using the extinction curve of Savage \& Mathis (1979).
WCFHS93 comment that their dereddening leaves a bump from
1900\,\AA\ to 2100\,\AA\ and a shallow trough from 2100 to 2300\,\AA.
We find the same difficulty here:
the slightly higher value of $E_{B-V}$ adopted here does slightly better
at removing the 2200\,\AA\ depression,
but at the expense of producing a bigger bump at 2000\,\AA.
The choice of $R$ makes little difference,
but lower values help to reduce the bump marginally.
The value $R = 3.0$ adopted here is slightly below the
`standard' value $R = 3.1$.

WCFHS93 fitted the continuum flux to a quadratic function of wavelength.
The simple quadratic form does impressively well in fitting the
dereddened spectrum over the full range 1600\,\AA\ to 3300\,\AA\ 
(cf.\ Figure~\ref{fluxfe} and WCFHS93 Figure~2).
However, the quadratic form does not fit perfectly,
and there remains a residual discrepancy which is not well fitted
by a low order polynomial,
and which may possibly result from imperfect modeling of the extinction,
especially around the 2200\,\AA\ bump.
The imperfection mainly affects the \ion{Fe}{2} 2383\,\AA\ feature:
the quadratic continuum appears too steep compared to the `true' continuum
around this feature.
Here we resolve the difficulty by the expedient of fitting
the dereddened continua around each of the two broad \ion{Fe}{2} features
separately, to two different linear functions in $\log F$-$\log \lambda$.
The adopted continua are shown in Figure~\ref{fluxfe}.

An important difference between the present analysis and that of
WCFHS93 is in the treatment of narrow lines.
WCFHS93's procedure was to identify all lines with an equivalent width,
defined relative to a local continuum,
greater than 3 times the expected error.
WCFHS93 then subtracted the best fitting Gaussian for each such line,
treating the position, width, and strength of each line as free parameters.
Here we adopt a different policy,
requiring that the positions, widths, and strengths of narrow
lines conform to prior expectation.
That is,
for each identified narrow line we subtract a Gaussian profile in which
the wavelength is set equal to the expected wavelength
(modulo an overall $+ 36 \,\kms$ shift for all lines),
the width is set equal to the instrumental resolution
(2.8\,\AA\ FWHM for G270H),
and the strengths are required to be mutually consistent
with other narrow lines of the same ion.

The relevant narrow lines are those in and near the
broad \ion{Fe}{2} features.
In the \ion{Fe}{2} 2383, 2344, 2374\,\AA\ feature,
besides the narrow (presumed interstellar) components of the \ion{Fe}{2}
lines themselves,
we identify the narrow line at 2298\,\AA\ 
as stellar \ion{C}{3} 2297.578\,\AA\ 
(Bruhweiler, Kondo \& McCluskey 1981).
WCFHS93 subtracted an unidentified narrow line at 2316\,\AA,
but the G190H spectrum does not confirm the reality of this line
in the G270H spectrum,
and here we leave it unsubtracted.
In the \ion{Fe}{2} 2600, 2587\,\AA\ feature,
besides the narrow \ion{Fe}{2} lines themselves,
we identify narrow lines of
\ion{Mn}{2} 2577, 2594, 2606\,\AA,
as did WCFHS93.

The mean velocity shift of the three most prominent narrow \ion{Fe}{2} lines,
those at 2383\,\AA, 2600\,\AA, and 2587\,\AA,
is $+ 36 \pm 24 \,\kms$, and we adopt this velocity shift for all the narrow
\ion{Fe}{2} and \ion{Mn}{2} lines.
We allow the stellar \ion{C}{3} 2298\,\AA\ line its own
best fit velocity shift of $+ 19 \,\kms$,
since there is no reason to assume that the stellar and interstellar
velocities coincide exactly.

The observed equivalent widths
of the \ion{Mn}{2} lines are approximately
proportional to their oscillator strengths times wavelengths,
which suggests the lines are unsaturated,
so we fix the ratios of the fitted \ion{Mn}{2} lines at
their unsaturated values.

Of the five narrow \ion{Fe}{2} lines,
the two principal lines
\ion{Fe}{2} 2383\,\AA\ and \ion{Fe}{2} 2600\,\AA,
and also the line with the fourth largest oscillator strength,
\ion{Fe}{2} 2587\,\AA,
have approximately equal observed equivalent widths (in velocity units)
relative to a local continuum,
$W_{2383}$, $W_{2600}$, $W_{2587}$ =
$85 \pm 6 \,\kms$, $64 \pm 6 \,\kms$, $62 \pm 6 \,\kms$
respectively
(at fixed centroid and dispersion),
which suggests the lines are saturated.
%We therefore fix the strengths of these lines to be equal.
The fifth and weakest line, \ion{Fe}{2} 2374\,\AA,
has an observed equivalent width about half that of the strong lines,
which is consistent with the weak line being marginally optically thin
and the strong lines again being saturated.
%We allow the strength of this line to take its best fit value.
For these four lines we allow the strength of the fitted line to take its
best fit value,
since they are mutually consistent within the uncertainties.

The line with the third largest oscillator strength,
\ion{Fe}{2} 2344\,\AA,
appears anomalous, since the observed line has an equivalent width
less than $1/4$ that of the strong lines,
or $1/2$ that of the intrinsically weaker \ion{Fe}{2} 2374\,\AA\ line.
In the fit,
we force the equivalent width of the anomalous \ion{Fe}{2} 2344\,\AA\ 
narrow line to be the saturated value measured from the
\ion{Fe}{2} 2600\,\AA\ and \ion{Fe}{2} 2587\,\AA\ lines,
multiplied by $0.8$ to allow for a $2 \sigma$ uncertainty in this value.
The fit gives the impression that the anomalous \ion{Fe}{2} 2344\,\AA\ line
is oversubtracted,
but the effect is to bring the profile of the deconvolved broad
\ion{Fe}{2} 2383\,\AA\ line into closer agreement with
that of \ion{Fe}{2} 2600\,\AA.

\begin{figure*}[tb]
\begin{minipage}{175mm}
\epsfbox[52 252 540 533]{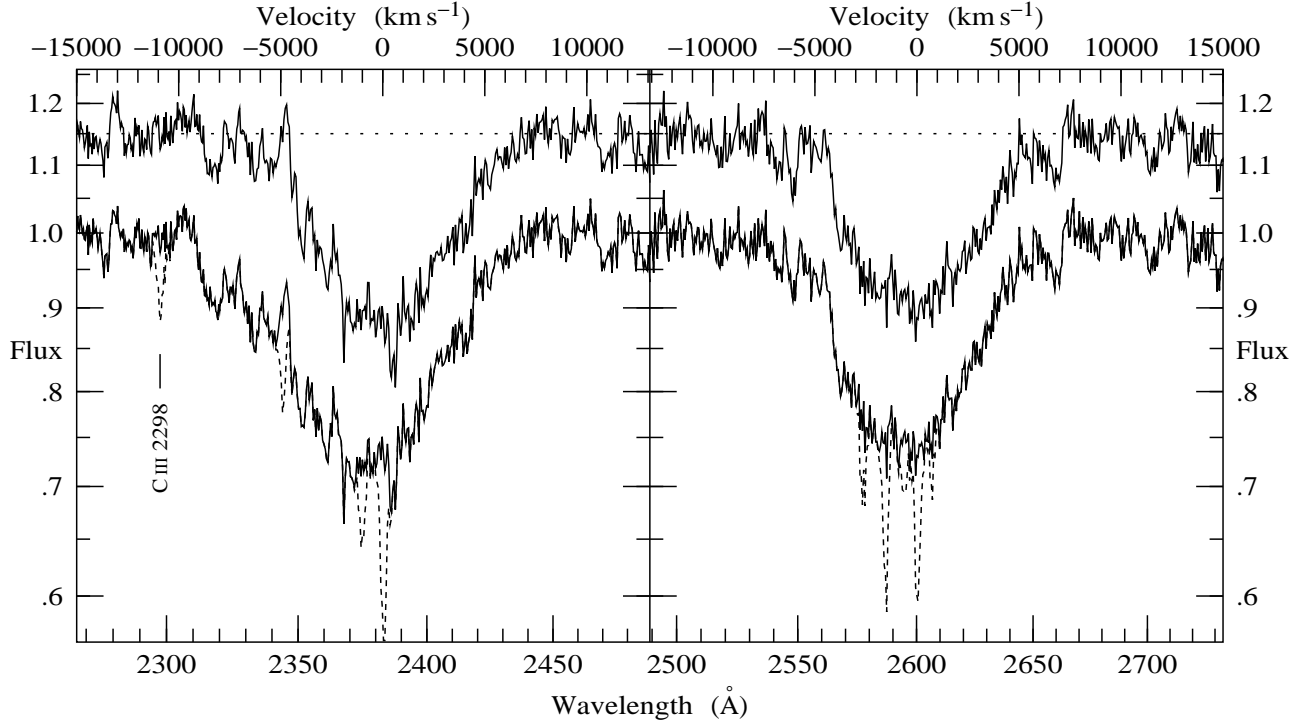}
%\vbox to98mm{\rule{0pt}{98mm}}
%\special{psfile=fe2.ps angle=0 hoffset=-57 voffset=-255 vscale=100 hscale=100}
  \caption[1]{
G270H spectrum showing at left the
\ion{Fe}{2} 2382.765, 2344.214, 2374.4612\,\AA\
($f = 0.3006$, 0.1097, 0.02818;
we ignore an unobserved fourth component \ion{Fe}{2} 2367.5905\,\AA\ 
with $f = 1.6 \times 10^{-4}$) feature,
and at right the
\ion{Fe}{2} 2600.1729, 2586.6500\,\AA\ 
($f = 0.2239$, 0.06457)
feature
(Morton 1991).
Below 2330\,\AA, the spectrum is an inverse-variance-weighted merger
of G190H and G270H spectra.
The lower curve shows the dereddened spectrum,
scaled to a continuum,
with narrow interstellar \ion{Fe}{2},
stellar \ion{C}{3} 2297.578\,\AA,
and interstellar \ion{Mn}{2} 2576.877, 2594.499, 2606.462\,\AA\ 
($f = 0.3508$, 0.2710, 0.1927)
lines subtracted
as indicated by dashed lines.
The upper curve (offset by $\log 1.15$)
shows the deconvolved spectra, after removal of
the weaker components of the broad \ion{Fe}{2} lines.
The velocity scales shown on the upper axis
are with respect to the rest frames of the principal component of each
of the features, the
\ion{Fe}{2} 2383\,\AA\ line on the left,
and the
\ion{Fe}{2} 2600\,\AA\ line on the right.
The adopted continua are as shown in Figure~\protect\ref{fluxfe}.
\label{fe2}
}
\end{minipage}
\end{figure*}

We deconvolved the broad
\ion{Fe}{2} 2383, 2344, 2374\,\AA\ and \ion{Fe}{2} 2600, 2587\,\AA\ 
features by subtracting the contributions from the weaker components,
using the following analytic procedure.
In a two component line,
the observed optical depth $\tau ( v )$
at velocity $v$ with respect to the principal component
is a sum of the line profile $\phi ( v )$ of the principal component
and the line profile $\epsilon \phi ( v + \Delta v )$ of the secondary
component, where
$\Delta v$ is the velocity shift of the secondary relative to the principal
component,
and $\epsilon = f_2 \lambda_2 / ( f_1 \lambda_1 ) < 1$
is the ratio of oscillator strengths times wavelengths:
\be
\label{tauv}
	\tau ( v ) = \phi ( v ) + \epsilon \phi ( v + \Delta v )
	\ .
\ee
Equation (\ref{tauv}) inverts to
\be
\label{phi}
	\phi ( v ) = \tau ( v ) - \epsilon \tau ( v + \Delta v )
	+ \epsilon^2 \tau ( v + 2 \Delta v )
%	- \epsilon^3 \tau ( v + 3 \Delta v )
	+ \cdots
\ee
which can conveniently be solved iteratively by
\ba
\label{phin}
	\phi_1 ( v ) &=& \tau ( v ) - \epsilon \tau ( v + \Delta v )
	\ , \nn
	\phi_{n+1} ( v )
	&=& \phi_n ( v ) + \epsilon^{2^n} \phi_n ( v + 2^n \Delta v )
	\ .
\ea
The iterative procedure converges rapidly to the solution,
$\phi_n \rightarrow \phi$, as $n$ increases;
we stop at $\phi_3$.
To avoid irrelevant parts of the spectrum outside the line profile
from propagating through the solution,
we set the optical depth to zero, $\tau ( v ) = 0$ in equation (\ref{phin}),
at velocities $v > 10000\,\kms$.
The above procedure works for a two component line such as
\ion{Fe}{2} 2600, 2587\,\AA,
and a slightly more complicated generalization works for a three component
line such as
\ion{Fe}{2} 2383, 2344, 2374\,\AA.

\subsection{\protect\ion{Fe}{2} line profiles}
\label{fe2sec}

Figure~\ref{fe2} shows the results of our reanalysis.
The upper curves in the Figure show the deconvolved 
\ion{Fe}{2} 2383\,\AA\ and \ion{Fe}{2} 2600\,\AA\ line profiles,
and these deconvolved profiles agree well with each other.
The deconvolved \ion{Fe}{2} 2600\,\AA\ profile here also agrees well with
that of WCFHS93.
However, the revised \ion{Fe}{2} 2383\,\AA\ profile
no longer shows compelling evidence
for high velocity blueshifted absorption beyond $- 4500 \,\kms$,
although the presence of some absorption is not excluded.

What causes the difference between the \ion{Fe}{2} 2383\,\AA\ line profile
shown in Figure~\ref{fe2} versus that of WCFHS93?
One factor is that we adopt different continua,
as illustrated in Figure~\ref{fluxfe}.
WCFHS93's single quadratic fit to the continuum over the entire spectrum
is certainly more elegant than the two separate linear fits
to each of the two broad \ion{Fe}{2} features which we adopt here.
The advantage of the fit here is that it removes the apparent tilt in the
\ion{Fe}{2} 2383\,\AA\ line profile left by the quadratic fit,
evident in Figures~2 and 3 of WCFHS93.

However, the major difference between the two analyses is
the subtraction here of the narrow \ion{Fe}{2} 2344\,\AA\ interstellar line
with a strength $0.8$ times the saturated line strength observed
in the \ion{Fe}{2} 2600\,\AA\ and 2587\,\AA\ narrow lines.
By comparison, WCFHS93 subtracted the narrow \ion{Fe}{2} 2344\,\AA\ line
using the observed strength of the line, which is anomalously weak compared
to the other four narrow \ion{Fe}{2} lines.

\begin{figure}[tb]
\epsfbox[162 270 410 498]{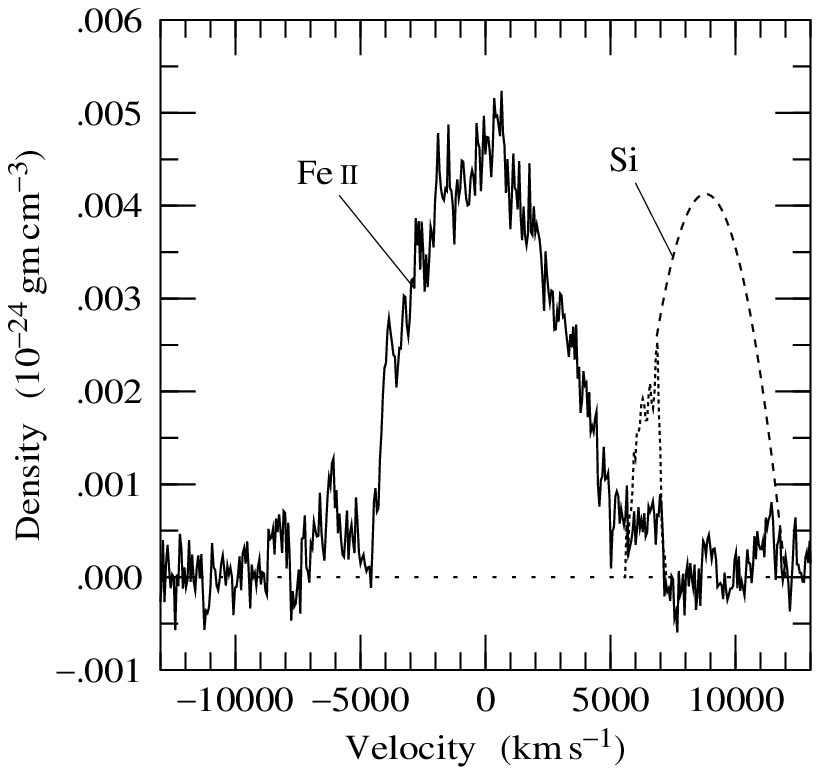}
%\vbox to80mm{\rule{0pt}{80mm}}
%\special{psfile=rho.ps angle=0 hoffset=-165 voffset=-272 vscale=100 hscale=100}
  \caption[1]{
Inferred density profile of ejecta in SN1006.
The \ion{Fe}{2} profile is from the mean of the two deconvolved
broad \ion{Fe}{2} absorption features shown in Figure~\protect\ref{fe2}.
%The Gaussian profile centered at $-2000 \,\kms$ illustrates
%the possible contribution to the absorption of shocked \ion{Fe}{2},
%as estimated in subsection~\protect\ref{shockedFe};
%the actual contribution of shocked \ion{Fe}{2} could be lower.
The dotted line at $+5600$-$7070 \,\kms$ is the profile of unshocked
\ion{Si}{2} from the redshifted \ion{Si}{2} 1260\,\AA\ absorption feature
in Figure~\protect\ref{si1260}.
The dashed line above $7070 \,\kms$ is a plausible extrapolation
of the total Si density before it was shocked:
it is a quadratic function of velocity, equation (\protect\ref{nSiquad}),
which approximately reproduces the observed profile of unshocked \ion{Si}{2},
and which contains the observed total Si mass of $0.25 \,\Msun$
(assuming spherical symmetry),
equation~(\protect\ref{MSi}).
\label{rho}
}
\end{figure}

%--------------------
% ApJ version
\begin{deluxetable}{lr}
\tablewidth{0pt}
\tablecaption{Parameters measured from \protect\ion{Fe}{2} profile
\label{fe2tab}}
\tablehead{\colhead{Parameter} & \colhead{Value}}
\startdata
Position of reverse shock on near side & $-4200 \pm 100 \,\kms$ \nl
Column density of \protect\ion{Fe}{2} &
$10.8 \pm 0.9 \times 10^{14} \,\cm^{-2}$ \nl
Mass of \protect\ion{Fe}{2} up to $7070\,\kms$ & $0.029 \pm 0.004 \,\Msun$ \nl
\enddata
\tablecomments{
\protect\ion{Fe}{2} mass is from red side of profile,
and assumes spherical symmetry.
}
\end{deluxetable}
%--------------------

%%--------------------
%% MNRAS version
%\begin{table}
%\caption{Parameters measured from \protect\ion{Fe}{2} profile
%\label{fe2tab}}
%\begin{tabular}{@{}lr}
%Parameter & \multicolumn{1}{c}{Value} \\ [4pt]
%Position of reverse shock on near side & $-4200 \pm 100 \,\kms$ \\
%Column density of \protect\ion{Fe}{2} &
%$10.8 \pm 0.9 \times 10^{14} \,\cm^{-2}$ \\
%Mass of \protect\ion{Fe}{2} up to $7070\,\kms$ &
%$0.029 \pm 0.004 \,\Msun$ \\ [5pt]
%\end{tabular}
%\protect\ion{Fe}{2} mass is from red side of profile,
%and assumes spherical symmetry.
%\end{table}
%%--------------------

Figure~\ref{rho} shows the \ion{Fe}{2} density inferred
from the mean of the two deconvolved \ion{Fe}{2} features.
The \ion{Fe}{2} column density inferred from the mean profile,
integrated from $- 4500 \,\kms$ to $+ 7100 \,\kms$,
is
\be
\label{NFeII}
	N_\FeII = 10.8 \pm 0.9 \times 10^{14} \,\cm^{-2}
	\ .
\ee
Most of the uncertainty,
based here on the scatter between the two deconvolved profiles,
comes from the blue side of the profile:
the column density on the blue side
from $- 4500 \,\kms$ to $0 \,\kms$ is
$N_\FeII = 5.2 \pm 0.8 \times 10^{14} \,\cm^{-2}$,
while the column density on the red side
from $0 \,\kms$ to $+ 7100 \,\kms$ is
$N_\FeII = 5.6 \pm 0.2 \times 10^{14} \,\cm^{-2}$.

In estimating the mass of \ion{Fe}{2}
from the density profile in Figure~\ref{rho},
we take into account the small correction which results from the fact that
the SM star is offset by $2'.45 \pm 0'.25$ southward from the projected center
of the $15'$ radius remnant
(Schweizer \& Middleditch 1980).
The offset corresponds to a free expansion velocity of
$v_\perp = 1300 \pm 250 \,\kms$
at the $1.8 \pm 0.3 \,\kpc$ distance of the remnant.
If spherical symmetry is assumed,
then the mass is an integral over the density $\rho(v)$
at line-of-sight velocity $v$:
\be
\label{M}
	M =
	M ( <\! v_\perp )
	+
	t^3 \! \int_{0}^{v_{\max}} \! \rho(v) 
	(v^2 + v_\perp^2)^{1/2} \, v \dd v
\ee
where $M ( <\! v_\perp )$ is the mass inside the free-expansion velocity
$v_\perp$.
At a constant central density of
$\rho_\FeII = 0.005 \times 10^{-24} \,\gm\,\cm^{-3}$,
the mass inside $v_\perp = 1300 \,\kms$ would be
$M_\FeII(<\! v_\perp) = 0.0007 \,\Msun$,
and the actual \ion{Fe}{2} mass is probably slightly higher,
given that the density is increasing mildly inward.
The masses given below, equation (\ref{MFeII}), include a fixed
$M_\FeII(<\! v_\perp) = 0.001 \,\Msun$.
The factor $(v^2 + v_\perp^2)^{1/2}$ rather than $v$ in equation (\ref{M})
increases the \ion{Fe}{2} masses by a further $0.002 \,\Msun$,
so the masses quoted in equation (\ref{MFeII}) are altogether $0.003 \,\Msun$
larger than they would be if no adjustment for the offset of the SM star
were applied.

The total mass of \ion{Fe}{2} inferred from the cleaner, red side
of the profile, assuming spherical symmetry, is then
\be
\label{MFeII}
	M_\FeII = \left\{
	\begin{array}{cl}
	0.0156 \pm 0.0009 \,\Msun & ( v \leq 4200 \,\kms ) \\
	0.0195 \pm 0.0013 \,\Msun & ( v \leq 5000 \,\kms ) \\
	0.029 \pm 0.004 \,\Msun & ( v \leq 7070 \,\kms ) \ .
	\end{array}
	\right.
\ee
The uncertainties here are based on the scatter between the two
deconvolved \ion{Fe}{2} profiles,
and do not included systematic uncertainties arising from placement
of the continuum,
which mostly affects the outer, high velocity parts of the profile.
WCFHS93 obtained
$M_\FeII = 0.014 \,\Msun$ from the red side of the mean \ion{Fe}{2} profile,
which is lower than the
$M_\FeII = 0.029 \,\Msun$ obtained here mainly because
of the different placement of the continuum
(see Figure~\ref{fluxfe}),
and to a small degree because of
the adjustment applied here for the offset of the SM star.

Figure~\ref{rho} also shows for comparison
the density of unshocked Si.
Below $7070 \,\kms$,
the Si density profile is just the unshocked profile inferred from
the \ion{Si}{2} 1260\,\AA\ absorption, Figure~\ref{si1260}.
Above $7070 \,\kms$,
the Si density is a plausible extrapolation which is consistent
with observational constraints:
it is a quadratic function of the free expansion velocity $v$
\ba
	n_\Si &=&
	0.00413 \times 10^{-24} \,\gm\,\cm^{-3}\,
	\nn
	& & \times \ 
	{(v - 5600 \,\kms) ( 12000 \,\kms - v ) \over (3200 \,\kms)^2}
\label{nSiquad}
\ea
which approximately reproduces the observed profile of unshocked \ion{Si}{2},
and which contains, on the assumption of spherical symmetry,
a total Si mass of $0.25 \,\Msun$,
in accordance with equation~(\ref{MSi}).

\subsection{Reverse shock on the near side}
\label{nearside}

If the reanalysis of the \ion{Fe}{2} lines here is accepted,
then it is natural to interpret the sharp blue edge on the
broad \ion{Fe}{2} lines at $-4200 \,\kms$
as representing the free expansion radius of the reverse shock on the
near side of SN1006.
This identification is not as convincing as the identification
of the sharp red edge on the \ion{Si}{2} 1260\,\AA\ feature
as representing the radius of the reverse shock on the far side
at $7070 \,\kms$.

\begin{figure}[tb]
\epsfbox[190 312 440 479]{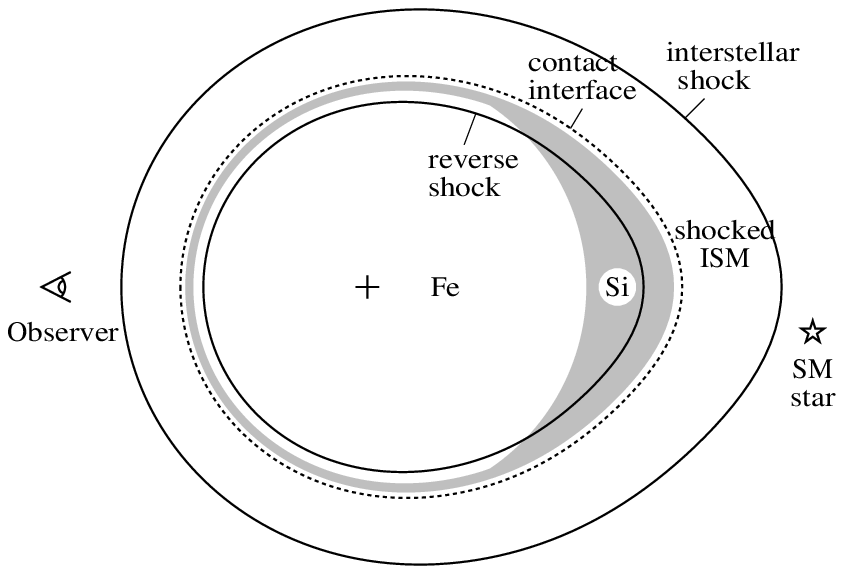}
%\vbox to62mm{\rule{0pt}{62mm}}
%\special{psfile=pic.ps angle=0 hoffset=-191 voffset=-312 vscale=100 hscale=100}
  \caption[1]{
Schematic diagram, approximately to scale,
of the structure of the remnant of SN1006 inferred in this paper.
The remnant on the far side bulges out because of the low interstellar
density there.
Shaded regions represent silicon ejecta, both shocked and unshocked.
Iron, both shocked and unshocked, lies inside the silicon.
The background SM star is offset slightly from the projected
center of the remnant.
\label{pic}
}
\end{figure}

Figure~\ref{pic} illustrates schematically
the inferred structure of the remnant of SN1006.
The picture is intended to be approximately to scale,
and in it the diameter of SN1006 along the line of sight
is roughly 20\% larger than the diameter transverse to the line of sight.
By comparison, the diameter of the observed radio and x-ray remnant
varies by 10\%,
from a minimum of $30'$ to a maximum of $33'$
(Reynolds \& Gilmore 1986, 1993)
or $34'$ (Willingale et al.\ 1996).

As already discussed in subsections~\ref{blue} and \ref{blueion},
if the position of the reverse shock on the near side at $4200 \,\kms$
is typical of the rest of the remnant,
while the $7070 \,\kms$ position of the reverse shock on the far side
is anomalously high because the interstellar density on the far side is low,
then it would resolve several observational puzzles.
In summary, these observational puzzles are:
(1) how to fit gas expanding at $\sim 7000 \,\kms$ within the confines
of the interstellar shock
(answer: the remnant bulges out on the far side because of the low density);
(2) how the $5050 \,\kms$ velocity of shocked Si on the far side could
be so much higher than the $1800 \,\kms$ velocity
(assuming no collisionless electron heating)
of gas behind the interstellar shock along the NW filament
(answer: velocities on the far side are anomalously high because
the interstellar density there is anomalously low);
(3) how to achieve Si densities $n_\Si \ga 10^{-2} \,\cm^{-2}$
necessary to produce the observed Si x-ray emission,
compared to the postshock Si density of $2.2 \times 10^{-4} \,\cm^{-2}$
measured from the \ion{Si}{2} absorption on the far side
(answer: gas shocked at earlier times is denser because density
decreases as $\rho \propto t^{-3}$ in free expansion);
and (4) why there is strong redshifted Si absorption but no blueshifted
absorption (answer: Si on the near side has been shocked and collisionally
ionized above \ion{Si}{4}).
As regards the second of these problems,
if the reverse shock on the near side is indeed at $4200 \,\kms$,
then the velocity of reverse-shocked gas on the near side
would be of order $2000 \,\kms$,
much more in keeping with the $1800 \,\kms$ velocity
of shocked gas in the NW filament.

\subsection{Contribution of shocked Fe to absorption}
\label{shockedFe}

In subsections~\ref{red} and \ref{SiIII+IV}
we concluded that most of the observed absorption by Si ions
is from shocked Si.
Does shocked Fe also contribute to the observed broad \ion{Fe}{2}
absorption profiles?
The answer, on both observational and theoretical grounds,
is probably not much.

On the red side of the \ion{Fe}{2} profile,
shocked \ion{Fe}{2} would have a Gaussian line profile
centered at $5050 \,\kms$, the same as observed for shocked Si.
No absorption with this profile is observed,
Figure~\ref{fe2} or \ref{rho}.
Since the collisional ionization rates of \ion{Fe}{2} and \ion{Si}{2}
are similar (Lennon et al.\ 1988),
the absence of such \ion{Fe}{2} absorption
implies that there cannot be much iron mixed in with the shocked Si.
This is consistent with the argument in subsection~\ref{purity},
that the bulk of the shocked \ion{Si}{2} must be fairly pure,
unmixed with other elements.

On the other hand the observed \ion{Fe}{2} profile, Figure~\ref{rho},
does suggest the presence of some \ion{Fe}{2} mixed with unshocked Si,
at velocities $\la 7070 \,\kms$.
The picture then is that there is Fe mixed with Si at lower velocities,
but not at higher velocities.
This is consistent with SN\,Ia models, such as W7 of Nomoto et al.\ (1984),
in which the transition from the inner Fe-rich layer to the Si-rich layer
is gradual rather than abrupt.

On the blue side of the \ion{Fe}{2} profile,
if the reverse shock is at $-4200 \,\kms$,
then shocked \ion{Fe}{2} should have a Gaussian profile centered at
$\sim - 2000 \,\kms$,
with a width comparable to that of the broad redshifted Si features.
While some such absorption may be present,
the similarity between the blueshifted and redshifted sides of the
\ion{Fe}{2} absorption suggests that the contribution
to the blue side from shocked \ion{Fe}{2} is not large.
This is consistent with expectation from
the density profile of unshocked \ion{Fe}{2} on the red side.
The mass of \ion{Fe}{2} at velocities $4200$-$7070 \,\kms$
inferred from the red side of the profile
on the assumption of spherical symmetry
is $0.013 \,\Msun$.
If this mass of \ion{Fe}{2} is supposed shocked on the blue side
and placed at the reverse shock radius of $- 4200 \,\kms$,
the resulting Fe column density is
$1.3 \times 10^{14} \,\cm^{-2}$,
which translates into a peak Fe density of
$0.0013 \times 10^{-24} \,\gm\,\cm^{-3}$
at velocity $-2200 \,\kms$,
for an assumed dispersion of $1240 \,\kms$
like that of the redshifted Si features.
This density of shocked Fe is low enough that it makes only
a minor contribution to the \ion{Fe}{2} profile in Figure~\ref{rho}.

In practice, collisional ionization of shocked \ion{Fe}{2}
reduces its contribution further.
For pure iron with an initial ionization state
of, say, 50\% \ion{Fe}{2}, 50\% \ion{Fe}{3}
(see subsection~\ref{photoion}),
we find that the column density of shocked \ion{Fe}{2} is reduced by
a factor 0.6 to
$0.8 \times 10^{14} \,\cm^{-2}$,
which translates into a peak \ion{Fe}{2} density of
$0.0008 \times 10^{-24} \,\gm\,\cm^{-3}$
at velocity $-2200 \,\kms$.
%The absorption profile of this amount of shocked \ion{Fe}{2}
%is shown in Figure~\ref{rho}.
The shocked column density would be even lower if the initial ionization
state is higher, or if there are other elements mixed in with the Fe,
since a higher electron to \ion{Fe}{2} ratio would
make collisional ionization faster.
If the initial ionization state of the iron is as high as proposed by HF88,
then the shocked \ion{Fe}{2} column density of Fe could be as low as
$0.1 \times 10^{14} \,\cm^{-2}$,
for a peak density of
$0.0001 \times 10^{-24} \,\gm\,\cm^{-3}$
at velocity $-2200 \,\kms$
in Figure~\ref{rho}.

\subsection{Photoionization}
\label{photoion}

The mass of \ion{Fe}{2} inferred here from the \ion{Fe}{2} profile is,
according to equation~(\ref{MFeII}),
$0.0195 \pm 0.0013 \,\Msun$ up to $5000 \,\kms$,
and $0.029 \pm 0.004 \,\Msun$ up to $7070 \,\kms$.

Historical and circumstantial evidence suggests that SN1006 was
a Type~Ia
(Minkowski 1966; Schaefer 1996).
Exploded white dwarf models of SN\,Ia
predict that several tenths of a solar mass of iron ejecta should be present,
as required to explain SN\,Ia light curves
(H\"{o}flich \& Khokhlov 1996).
Thus,
as emphasised by Fesen et al. (1988) and by HF88,
the observed mass of \ion{Fe}{2} in SN1006 is only a fraction
($\la 1/10$)
of the expected total Fe mass.

Hamilton \& Sarazin (1984)
pointed out that unshocked SN ejecta will be subject to photoionization
by UV starlight and by UV and x-ray emission from the reverse shock.
Recombination is negligible at the low densities here.
HF88 presented detailed calculations of the time-dependent
photoionization of unshocked ejecta in SN1006,
using deflagrated white dwarf models W7 of Nomoto et al.\ (1984),
and CDTG7 of Woosley (1987, private communication),
evolved by hydrodynamic simulation into a uniform interstellar medium.
HF88 found that in these models most of the unshocked Fe was photoionized
to \ion{Fe}{3}, \ion{Fe}{4}, and \ion{Fe}{5}.
While model W7 produced considerably more \ion{Fe}{2} than observed in SN1006,
model CDTG7, which is less centrally concentrated than W7,
produced an \ion{Fe}{2} profile in excellent agreement with the IUE
\ion{Fe}{2} 2600\,\AA\ feature.
HF88 concluded that several tenths of a solar mass of unshocked Fe
could be present at velocities $\la 5000 \,\kms$ in SN1006,
as predicted by Type~Ia supernova models.

However,
the low ionization state of unshocked Si
inferred from the present HST observations
does not support the high ionization state of Fe advocated by HF88.
According to the ion fractions given in Table~\ref{sitab},
unshocked Si is $92 \pm 7\%$ \ion{Si}{2}.
By comparison,
HF88 argued that unshocked Fe is only $\sim 10\%$ \ion{Fe}{2}.
We now show that these ionization states of \ion{Si}{2} and \ion{Fe}{2}
are not mutually consistent.

The ionizations of unshocked \ion{Si}{2} and \ion{Fe}{2} are related
according to their relative photoionization cross-sections,
and by the energy distribution of the photoionizing radiation.
Neutral Si and Fe can be neglected here,
since they have ionization potentials below the Lyman limit,
and are quickly ionized by UV starlight
(\ion{Si}{1} in $\sim 20 \,\yr$, \ion{Fe}{1} in $\sim 100 \,\yr$
if the starlight is comparable to that in the solar neighborhood),
once the ejecta start to become optically thin to photoionizing radiation,
at $\sim 100$-$200 \,\yr$.

\begin{figure}[tb]
\epsfbox[173 272 423 498]{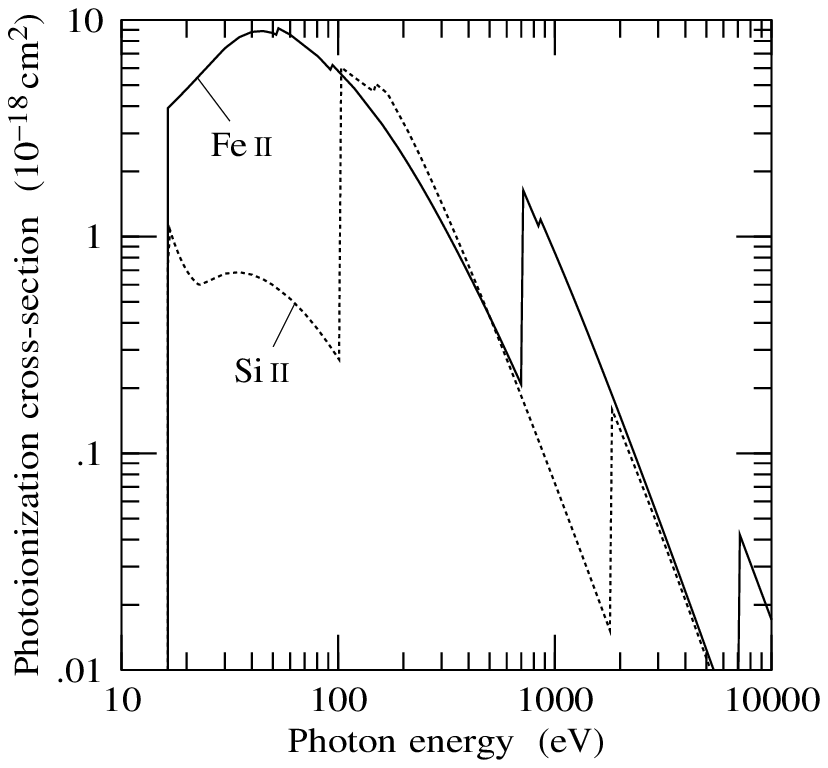}
%\vbox to80mm{\rule{0pt}{80mm}}
%\special{psfile=anu.ps angle=0 hoffset=-178 voffset=-272 vscale=100 hscale=100}
  \caption[1]{
Photoionization cross-sections of \ion{Si}{2} and \ion{Fe}{2}
(Reilman \& Manson 1979, adapted to include autoionizing photoionization
as described by HF88).
The ionization state of unshocked ejecta depends on photoionization,
and the plotted photoionization cross-sections are important in relating
the ionization state of unshocked iron,
characterized by the ratio \ion{Fe}{2}/Fe,
to the ionization state of unshocked silicon,
characterized by the ratio \ion{Si}{2}/Si.
\label{anu}
}
\end{figure}

Photoionization cross-sections of \ion{Si}{2} and \ion{Fe}{2},
taken from Reilman \& Manson (1979) and adapted to include
autoionizing photoionization as described by HF88,
are shown in Figure~\ref{anu}.
The Figure shows that
the photoionization cross-section of \ion{Fe}{2}
is about an order of magnitude larger than that of \ion{Si}{2}
from the ionization potential up to the L-shell (autoionizing) photoionization
threshold of \ion{Si}{2} at 100\,eV,
above which the photoionization cross-sections are about equal,
until the L-shell threshold of \ion{Fe}{2} at 700\,eV.
According to HF88 (Table~6, together with Table~2 of Hamilton \& Sarazin 1984),
much of the photoionizing emission from the reverse shock is in the UV
below 100\,eV.
However, there is also some soft x-ray emission above 100\,eV,
which is important for \ion{Si}{2} because
its L-shell photoionization cross-section is larger than
that of the valence shell.
Averaging over the photoionizing photons tabulated by HF88,
we find that the effective photoionization cross-section of \ion{Fe}{2}
is about 5 times that of \ion{Si}{2},
which is true whether the source of the emission in the reverse shock
is oxygen, silicon, or iron.

If the effective photoionization cross-section of \ion{Fe}{2}
is 5 times that of \ion{Si}{2},
then an unshocked \ion{Si}{2} fraction of $0.92 \pm 0.07$
would predict an unshocked \ion{Fe}{2} fraction of
$( 0.92 \pm 0.07 )^5 = 0.66^{+ 0.29}_{- 0.22}$,
considerably larger than the desired unshocked \ion{Fe}{2} fraction
of $0.1$.
The $3 \sigma$ lower limit on the \ion{Si}{2} fraction is $0.71$,
which would predict an unshocked \ion{Fe}{2} fraction of
$( 0.71 )^5 = 0.18$,
closer to but still higher than desired.
 
A higher ionization state of Fe compared to Si might be achieved if
the photoionizing emission could be concentrated entirely in the UV
below 100\,eV,
since then the effective photoionization cross-section of \ion{Fe}{2}
would be about 10 times that of \ion{Si}{2},
as illustrated in Figure~\ref{anu}.
This could occur if the photoionizing emission were mainly from He.
In this case,
the predicted unshocked \ion{Fe}{2} fraction would be
$( 0.92 \pm 0.07 )^{10} = 0.43^{+ 0.47}_{- 0.27}$,
with a $3 \sigma$ lower limit of
$( 0.71 )^{10} = 0.03$,
which is satisfactory.
To achieve this relatively high level of Fe ionization
requires that there be little soft x-ray emission from heavy elements.
It is not clear whether this is possible,
given the observed x-ray emission from oxygen and silicon
(Koyama et al.\ 1995).

Thus the low ionization state of unshocked Si inferred
in this paper is difficult to reconcile with the expected presence of
several tenths of a solar mass of Fe at velocities $\la 5000 \,\kms$.
Is it possible that the unshocked Si is substantially more ionized
than we have inferred?
From Tables~\ref{redtab} and \ref{sitab},
the total column densities of Si ions, shocked and unshocked, are in the ratio
\be
\label{NSitot}
	N_\SiII : N_\SiIII : N_\SiIV =
	1 : 0.39 : 0.36
	\ .
\ee
If the ionization state of the unshocked Si were this high,
then a high ionization state of Fe would be allowed,
and the problem would be resolved.
In fact, in photoionization trials similar to those described by HF88,
we find that the unshocked Si fractions predicted by the CDTG7 model
are close to the ratio (\ref{NSitot}).

However,
in subsection~\ref{SiIII+IV}
we argued both theoretically and from the observed line profiles
that most of the observed \ion{Si}{3} and \ion{Si}{4} absorption
is from shocked Si.
Might this be wrong,
and could in fact much or most of the absorption be from unshocked Si?
And is then our interpretation of the \ion{Si}{2} 1260\,\AA\ profile
as arising mostly from shocked \ion{Si}{2} also faulty?
If so, then much of the tapestry of reasoning in this paper begins to
unravel.
For example,
we must regard as merely coincidental the agreement,
found in subsection~\ref{jump},
of the measured parameters of the \ion{Si}{2} 1260\,\AA\ profile
with the energy shock jump condition, equations~(\ref{Dv})-(\ref{Dvobs}).
We must also conclude that the observed asymmetry between the red and
blueshifted Si absorption arises from asymmetry in the initial explosion,
not (or not all) from asymmetry in the ambient ISM.
For if the blue edge of the redshifted
\ion{Si}{2} and \ion{Si}{4} features
(the blue edge of redshifted \ion{Si}{3} is obscured by Ly\,$\alpha$)
arises from unshocked Si extending down to velocities $+ 2500 \,\kms$
(see Figs.~\ref{si1260} and \ref{si4}),
then there should be, on the assumption of spherical symmetry,
corresponding blueshifted Si absorption outward of $- 2500 \,\kms$,
which is not seen.

\subsection{Where's the iron?}
\label{where}

H\"{o}flich \& Khokhlov's (1996) Table~1 presents a survey of 37 SN\,Ia models,
encompassing all currently discussed explosion scenarios.
In their models, the ejected mass of $^{56}\Ni$,
which decays radioactively to iron,
ranges from $0.10 \,\Msun$ to $1.07 \,\Msun$.
Models yielding `acceptable' fits to the sample of 26 SN\,Ia considered by
H\"{o}flich \& Khokhlov
have ejected $^{56}\Ni$ masses of
$0.49$-$0.83 \,\Msun$ for normal SN\,Ia,
and $0.10$-$0.18 \,\Msun$ for subluminous SN\,Ia.
In the subluminous models,
a comparable amount of Fe is ejected along with the $^{56}\Ni$
(H\"{o}flich, Khokhlov \& Wheeler 1995),
so the total iron ejected in these cases is $\approx 0.2$-$0.3 \,\Msun$.

In the previous subsection,
we argued that the low ionization state of unshocked Si
inferred from the present observations suggests
that the ionization state of unshocked Fe is also likely to be low.
Specifically, if the unshocked \ion{Si}{2} fraction is
$\mbox{\ion{Si}{2}/Si} = 0.92 \pm 0.07$,
from Table~\ref{redtab},
then the predicted unshocked \ion{Fe}{2} fraction is
$\mbox{\ion{Fe}{2}/Fe} = 0.66^{+ 0.29}_{- 0.22}$.
Correcting the \ion{Fe}{2} mass of
$M_\FeII = 0.029 \pm 0.004 \,\Msun$ up to $7070 \,\kms$,
equation~(\ref{MFeII}),
for the ionization state of the Fe
yields a total inferred Fe mass of
$M_\Fe = 0.044^{+ 0.022}_{- 0.013} \,\Msun$
up to $7070 \,\kms$,
with a $3 \sigma$ upper limit of $M_\Fe < 0.16 \,\Msun$.
These Fe masses are lower than predicted by either normal or
subluminous models of SN\,Ia.

A low ionization state of Fe is supported by the HUT observations of
Blair et al.\ (1996),
who looked for \ion{Fe}{3} 1123\,\AA\ absorption in the background SM star.
If there is a large mass of Fe in SN1006,
significantly larger than the observed \ion{Fe}{2} mass,
then certainly \ion{Fe}{3} should be more abundant than \ion{Fe}{2},
and from detailed models HF88 predicted
$\mbox{\ion{Fe}{3}/\ion{Fe}{2}} = 2.6$.
Blair et al.'s best fit is 
$\mbox{\ion{Fe}{3}/\ion{Fe}{2}} = 1.1 \pm 0.9$,
and their $3 \sigma$ upper limit is
$\mbox{\ion{Fe}{3}/\ion{Fe}{2}} < 3.8$.
This result does not support, though it does not yet definitely exclude,
HF88's prediction.

Neither of the above two observational evidences favoring a low ionization
state of Fe, hence a low mass of Fe in SN1006, is yet definitive.
To settle the issue will require re-observation of the \ion{Fe}{3} 1123\,\AA\
line at a higher signal to noise ratio.
The Far Ultraviolet Space Explorer (FUSE) should accomplish this.

\section{Worries}
\label{worries}

In this paper we have attempted to present a consistent
theoretical interpretation of the broad Si and Fe absorption features
in SN1006.
While the overall picture appears to fit together nicely,
the pieces of the jigsaw do not fit perfectly everywhere.
In this Section we highlight the ill fits.

What causes the discrepancy between the profiles of the redshifted
\ion{Si}{2} 1260\,\AA\ and \ion{Si}{2} 1527\,\AA\ features?
This discrepancy was originally pointed out and discussed for the IUE data
by Fesen et al.\ (1988),
and the discrepancy remains in the HST data (WCHFLS96).
The discrepancy is especially worrying for the present paper
because the excess in the \ion{Si}{2} 1260\,\AA\ profile
compared to \ion{Si}{2} 1527\,\AA\ (see WCHFLS96, Figure~2)
looks a lot like what we
have interpreted here as the unshocked component of the
\ion{Si}{2} 1260\,\AA\ absorption (Figure~\ref{si1260}).

We have argued that the redshifted
\ion{Si}{2}, \ion{Si}{3}, and \ion{Si}{4}
absorption is caused mostly by shocked Si,
yet it is not clear that the observed relative column densities
are naturally reproduced in collisionally ionized gas.
Specifically, one might expect relatively more \ion{Si}{3},
or relatively less \ion{Si}{2} or \ion{Si}{4} in the shocked Si.
On the other hand the observed relative column densities of Si
are naturally reproduced in unshocked, photoionized gas.
Is our interpretation at fault?

The best fit dispersion of the redshifted \ion{Si}{4} feature,
Figure~\ref{si4},
is $1700 \pm 100 \,\kms$, which is $4.5 \sigma$ larger than that
of the \ion{Si}{2} 1260\,\AA\ feature, $1240 \pm 40 \,\kms$.
What causes this discrepancy?

Does the density of shocked Si vary a little or a lot?
In subsection~\ref{red} we argued that the Gaussian profile of the
shocked Si suggests little temperature variation,
hence little density variation.
On the other hand the unshocked Si density profile below $7070 \,\kms$
(Fig.~\ref{si1260})
suggests a density profile increasing steeply outwards,
and there are other clues hinting at the same thing:
the large column density of shocked \ion{Si}{2}, subsection~\ref{purity},
and the need for a high density to ionize Si on the near side above
\ion{Si}{4}, subsection~\ref{blueion}.
Is there an inconsistency here?

In subsection~\ref{purity} we argued that the high observed column density
of shocked \ion{Si}{2} indicates a low mean electron to \ion{Si}{2} ratio,
$n_e / n_\SiII \la 1.3$, equation (\ref{nela}).
Higher ratios would cause more rapid collisional ionization of \ion{Si}{2},
reducing the column density below what is observed.
This limit on the electron to \ion{Si}{2} ratio is
satisfactory as it stands
(the limit is somewhat soft, and at least it is not less than 1),
but it is uncomfortably low,
making it difficult to admit even modest quantities of other elements,
such as sulfur, which might be expected to be mixed with the silicon.
Is there a problem here,
and if so have we perhaps overestimated the contribution of shocked
compared to unshocked Si in the \ion{Si}{2} absorption?

In subsection~\ref{blueion}
we estimated that if the reverse shock on the near side of SN1006
is at $- 4200 \,\kms$,
then under a `simplest' set of assumptions
there should be an observable column density of blueshifted \ion{Si}{4},
contrary to observation.
However, we also showed that the predicted column density is sensitive
to the assumptions, and that it is not difficult to bring the column
density of \ion{Si}{4} below observable levels.
Is this explanation adequate,
or does the absence of blueshifted Si absorption
hint at asymmetry in the supernova explosion?

Is the sharp blue edge on the \ion{Fe}{2} features at $- 4200 \,\kms$ real?
Further observations of the \ion{Fe}{2} features at higher resolution
would be helpful in deciding this issue.

Notwithstanding our reanalysis of the \ion{Fe}{2} features,
there remains some suggestion of high velocity blueshifted absorption
outside $- 4200 \,\kms$, perhaps to $- 7000 \,\kms$ or even farther,
Figures~\ref{fe2} and \ref{rho}.
Is this absorption real?
If so,
then the arguments of subsection~\ref{blue} fail,
and the absence of blueshifted Si absorption must be attributed
to intrinsic asymmetry in the initial supernova explosion.

Finally, there is the problem discussed in subsection~\ref{where}:
where is the iron?

\section{Summary}
\label{summary}

We have presented a consistent
interpretation of the broad Si and Fe absorption features
observed in SN1006 against the background SM star
(Schweizer \& Middleditch 1980).

We have argued that the strong redshifted \ion{Si}{2} 1260\,\AA\ absorption
feature arises from both unshocked and shocked Si,
with the sharp red edge of the feature at $7070 \,\kms$ representing
the free expansion radius of the reverse shock on the far side of SN1006,
and the Gaussian blue edge signifying shocked Si
(Fig.~\ref{si1260}).
Fitting to the \ion{Si}{2} 1260\,\AA\ line profile yields three velocities,
the position of the reverse shock,
and the velocity and dispersion of the shocked gas,
permitting a test of the energy jump condition for a strong shock.
The measured velocities satisfy the condition remarkably well,
equations~(\ref{Dv})-(\ref{Dvobs}).
The \ion{Si}{2} 1260\,\AA\ line thus provides direct evidence
for the existence of a strong shock under highly collisionless conditions.

The energy jump condition is satisfied provided that virtually all the shock
energy goes into ions.
This evidence suggests little or no collisionless heating of electrons
in the shock,
in agreement with recent evidence from UV line widths and strengths
(Raymond et al.\ 1995; Laming et al.\ 1996).

The observed column density of shocked \ion{Si}{2} is close to the column
density expected for steady state collisional ionization behind a shock,
provided that the electron to \ion{Si}{2} ion ratio is low.
From the low electron to \ion{Si}{2} ratio, we have argued
that the shocked Si is probably of a fairly high degree of purity,
with little admixture of other elements.
More directly,
the absence of \ion{Fe}{2} absorption with the same line profile as
the shocked Si indicates that there is little Fe mixed with the shocked Si.
On the other hand,
there is some indication of absorption by \ion{Fe}{2}
at the velocity 5600-$7070 \,\kms$ of the unshocked \ion{Si}{2},
which suggests that some Fe is mixed with Si in the lower velocity region
of the Si layer.

We have proposed that the ambient interstellar density on the far side
of SN1006 is anomalously low compared to the density around the rest of
the remnant,
so that the remnant bulges out on the far side (Fig.~\ref{pic}).
This would explain several observational puzzles.
Firstly,
it would explain the absence of blueshifted Si absorption
matching the observed redshifted Si absorption.
If the interstellar density on the near side is substantially larger than
on the far side,
then the reverse shock on the near side would be further in,
so that all the Si on the near side could have been shocked and collisionally
ionized above \ion{Si}{4}, making it unobservable in absorption.
Secondly,
if the velocity on the far side is anomalously high because of the low
interstellar density there,
it would resolve the problem noted by Wu et al.\ (1983) and
subsequent authors of how to fit gas expanding at $\sim 7000 \,\kms$
within the confines of the interstellar shock.
Thirdly,
a low density on the far side
would explain how the $5050 \,\kms$ velocity of shocked Si there
could be so much higher than the $1800 \,\kms$ velocity
(assuming no collisionless electron heating)
of gas behind the interstellar shock along the NW filament
(Smith et al.\ 1991;
Raymond et al.\ 1995).
Finally,
the density of Si on the far side inferred from the Si absorption profiles
is one or two orders of magnitude too low to yield Si x-ray emission
at the observed level
(Koyama et al.\ 1995).
Again, an anomalously low density on the far side is indicated.

The notion that the reverse shock on the near side
has moved inward much farther, to lower velocities, than on the far side
conflicts with our earlier conclusion (WCFHS93)
that there is blueshifted \ion{Fe}{2} absorption to velocities
$\sim - 8000 \,\kms$.
Reanalyzing the \ion{Fe}{2} data,
we find that the evidence for such high velocity blueshifted \ion{Fe}{2}
absorption is not compelling.
In the WCFHS93 analysis,
the main evidence for high velocity blueshifted \ion{Fe}{2} comes from the
\ion{Fe}{2} 2383, 2344, 2374\,\AA\ feature.
However, the 2344\,\AA\ component, which lies at $- 4900 \,\kms$
relative to the principal 2383\,\AA\ component,
confuses interpretation of the blue wing of the feature.
The \ion{Fe}{2} 2600, 2587\,\AA\ feature is cleaner,
and it shows a sharp blue edge at $- 4200 \,\kms$,
which we interpret as representing the free expansion radius of the
reverse shock on the near side of SN1006.
In our reanalysis of the \ion{Fe}{2} features,
we adopt a rigorous approach to the subtraction of narrow
interstellar and stellar lines,
requiring that lines subtracted have the correct positions and dispersions,
and have mutually consistent strengths.
In particular,
we subtract the narrow \ion{Fe}{2} 2344\,\AA\ line with a strength
consistent with the other narrow \ion{Fe}{2} lines,
which strength is substantially greater than the apparent strength.
Subtraction of this line introduces a sharp blue edge on the
deconvolved \ion{Fe}{2} 2383, 2344, 2374\,\AA\ feature
at the same place, $\approx - 4200 \,\kms$,
as the \ion{Fe}{2} 2600, 2587\,\AA\ feature.
The resulting deconvolved \ion{Fe}{2} profiles
(Fig.~\ref{fe2})
are in good agreement with each other.

The mass and velocity distribution of Si and Fe inferred in this paper
provides useful information for modeling the remnant of SN1006
(see Fig.~\ref{rho}).
Freely expanding unshocked Si on the far side extends from a low velocity of
$5600 \pm 100 \,\kms$ up to the position of the reverse shock at $7070 \,\kms$.
Above this velocity the Si is shocked,
and information about its detailed velocity distribution
before being shocked is lost.
The total mass of Si, both unshocked and shocked,
inferred from the \ion{Si}{2}, \ion{Si}{3}, and \ion{Si}{4} lines
is $M_\Si = 0.25 \pm 0.01 \,\Msun$,
on the assumption of spherical symmetry.

We have argued that the observed broad \ion{Fe}{2} absorption arises
almost entirely from unshocked freely expanding Fe.
The mass of \ion{Fe}{2} inferred from the cleaner, red side of the
mean \ion{Fe}{2} profile is
$M_\FeII = 0.0195 \pm 0.0013 \,\Msun$ up to $5000 \,\kms$,
and $M_\FeII = 0.029 \pm 0.004 \,\Msun$ up to $7070 \,\kms$,
again on the assumption of spherical symmetry.
These masses include a small positive adjustment ($0.003 \,\Msun$)
resulting from
the offset of the SM star from the projected center of the remnant.

Our analysis of the Si lines indicates a low ionization state for
the unshocked silicon, with \ion{Si}{2}/Si = $0.92 \pm 0.07$.
Such a low state would imply a correspondingly low ionization state of
unshocked iron,
with \ion{Fe}{2}/Fe = $0.66^{+ 0.29}_{- 0.22}$.
If this is correct,
then the total mass of Fe up to $7070 \,\kms$ is
$M_\Fe = 0.044^{+ 0.022}_{- 0.013} \,\Msun$
with a $3 \sigma$ upper limit of $M_\Fe < 0.16 \,\Msun$.
The absence of \ion{Fe}{2} absorption with a profile like that of
the shocked \ion{Si}{2} suggests that there is not much more Fe
at higher velocities.
Such a low mass of Fe conflicts with the expectation that there should
be several tenths of a solar mass of Fe in this suspected Type~Ia remnant.
A low ionization state of Fe and a correspondingly low Fe mass
is consistent with the low
\ion{Fe}{3}/\ion{Fe}{2} $= 1.1 \pm 0.9$
ratio measured by Blair et al.\ (1996) from HUT observations of the
\ion{Fe}{3} 1123\,\AA\ line in the spectrum of the SM star.
However,
neither the present observations nor the HUT data are yet conclusive.

Re-observation of the \ion{Fe}{3} 1123\,\AA\ line at higher signal to noise
ratio with FUSE will be important in determining the ionization state of
unshocked Fe in SN1006,
and in resolving the question, Where's the iron?

%--------------------
% ApJ version
\acknowledgements
%--------------------

%%--------------------
%% MNRAS version
%\bigskip
%\noindent{\bf ACKNOWLEDGMENTS}
%\bib\strut
%
%\noindent
%%--------------------

We would like to thank Bill Blair for helpful correspondence on the HUT data,
and Graham Parker and Mike Shull for advice
on respectively stellar and interstellar lines.
Support for this work was provided by NASA through grant number
GO-3621
from the Space Telescope Science Institute,
which is operated by AURA, Inc., under NASA contract NAS 5-26555.

%--------------------
% ApJ version
\appendix
\section{Ionization times in non-steady state}
%--------------------

%%--------------------
%% MNRAS version
%\section*{APPENDIX}
%%--------------------

This Appendix demonstrates the assertion made in subsection~\ref{purity},
that the column density per interval of ionization time
is reduced compared to the steady state value if the density of shocked gas
increases downstream, and vice versa.
The main result is equation (\ref{dNdtau}).

Define a collisional ionization time $\tau$
(not to be confused with optical depth)
by the integral over time of the number density $n$ in a Lagrangian
gas element, from the time $t_s$ when the gas element was first shocked,
to the present time $t$:
\be
\label{tauion}
	\tau \equiv \int_{t_s}^t n\,\dd t
	\ .
\ee
The density $n$ in the definition (\ref{tauion}) of the ionization time
is the ion density.
One might think to use the electron density $n_e$ rather than the ion density,
since the collisional ionization is by electron impact,
but in fact the electron to ion ratio is itself determined by the ionization
time, for any given initial ionization state,
so there is no loss of generality in the definition (\ref{tauion}).

It is useful to regard the shock age $t_s$ as a Lagrangian coordinate
labeling different gas elements.
Then the ionization time $\tau$ of different gas elements at fixed time $t$
varies according to
\be
\label{dtau}
	\left. {\partial \tau \over \partial t_s} \right|_t
	= - n_s + \int_{t_s}^t
	\left. {\partial n \over \partial t_s} \right|_t \,\dd t
\ee
where $n_s$ is the postshock density at the time the gas was shocked.
The minus sign in front of $n_s$ reflects the fact that
gas elements farther downstream were shocked at earlier times.
If the pressure is spatially constant, which should be a good approximation,
and if the gas density varies adiabatically as the pressure changes in time,
then the densities $n$ in different elements vary with time
in proportion to each other.
It follows that the logarithmic derivative of density $n$ with respect to
the Lagrangian coordinate $t_s$ is independent of time.
It is convenient to denote this dimensionless logarithmic derivative by
$\gamma$:
\be
\label{gamma}
	\gamma ( t_s ) \equiv
	- \left. {\partial \ln n \over \partial \ln t_s} \right|_t
\ee
which is positive if the density at fixed time $t$ increases downstream.
Then equation (\ref{dtau}) becomes
\be
\label{dtau'}
	\left. {\partial \tau \over \partial \ln t_s} \right|_t
	= - n_s t_s - \gamma \tau
	\ .
\ee

Now consider the column density $N$ contributed by gas elements shocked
at different times $t_s$.
The rate at which mass $M$ of ions mass $m$ enters the shock at time $t_s$,
when the shock radius is $r_s$, is
\be
	\dot M
	= - {\dd M \over \dd t_s}
	= 4 \pi m r_s^2 n_s^{\rmn presh} v_s
	= \pi m r_s^2 n_s v_s
\ee
the postshock density being 4 times the preshock density for a strong shock,
$n_s = 4 n_s^{\rmn presh}$.
The minus sign $- \dd M / \dd t_s$ here arises because we choose
to label the interior mass $M ( t_s )$ so that it increases downstream
of the shock, in the direction of decreasing $t_s$.
Mass is conserved in Lagrangian elements,
and the column density $N$ at time $t$ when the shock radius is $r$
varies with $t_s$ as
\be
\label{dN}
	\left. {\partial N \over \partial \ln t_s} \right|_t
	= {1 \over 4 \pi m r^2} {\dd M \over \dd \ln t_s}
	= - {r_s^2 \over r^2} {n_s v_s t_s \over 4}
\ee
which is diluted by the geometric factor $r_s^2 / r^2$
compared to the column density at the time $t_s$ when the gas was shocked.

Comparing the column density (\ref{dN}) to the ionization time (\ref{dtau'})
shows that the column density per interval of ionization time
observed at a time $t$ is
\be
\label{dNdtau}
	\left. {\partial N \over \partial \tau} \right|_t
	= {r_s^2 v_s \over 4 r^2} {1 \over [ 1 + \gamma \tau / (n_s t_s ) ]}
\ee
which is the principal result of this Appendix.
In steady state, equation (\ref{dNdtau}) reduces to
$\partial N / \partial \tau |_t = v_s / 4$.
Compared to steady state,
the non-steady state column density per unit ionization time
(\ref{dNdtau})
thus contains two factors discussed in subsection~\ref{purity},
the geometric dilution factor $r_s^2/r^2$,
and the density profile factor $[1 + \gamma \tau / (n_s t_s)]^{-1}$.
Since $\tau$ and $n_s t_s$ are always positive,
this latter factor is less than or greater than one depending on whether
$\gamma$, equation (\ref{gamma}), is greater than or less than zero.
In other words,
if the density $n$ increases downstream, $\gamma > 0$,
then the column density is reduced compared to steady state,
and vice versa, as was to be proved.

\references
%--------------------

%%--------------------
%% MNRAS version
%\bigskip
%\noindent{\bf REFERENCES}
%\bib \strut
%%--------------------

\bib Arnett D., \& Livne E., 1994, ApJ, 427, 330

\bib Becker R. H., Szymkowiak A. E., Boldt E. A., Holt S. S.,
\& Serlemitsos P. J., 1980, ApJ, 240, L33

\bib Blair W. P., Long K. S., \& Raymond J. C. 1996, ApJ, 468, 871

\bib Bruhweiler F. C., Kondo Y., \& McCluskey G. E. 1981, ApJS, 46, 255

\bib Cardelli J. A., Clayton G. C., \& Mathis J. S. 1989, ApJ, 345, 245

\bib Chevalier R. A., Blondin J. M., \& Emmering R. T. 1992, ApJ, 392, 118

\bib Fesen R. A., \& Hamilton A. J. S. 1988,
in A Decade of UV Astronomy with IUE\@.  ESA SP-281, Vol.\ 1, p.\ 121

\bib Fesen R. A., Wu C.-C., Leventhal M., \& Hamilton A. J. S. 1988,
ApJ 327, 164

\bib Galas C. M. F., Venkatesan D., \& Garmire G. 1982, ApJ, 22, 103

\bib Garcia-Senz D., \& Woosley S. E. 1995, ApJ, 454, 895

\bib Hamilton A. J. S., \& Fesen R. A. 1988, ApJ, 327, 178 (HF88)

\bib Hamilton A. J. S., \& Sarazin C. L. 1984, ApJ, 287, 282

\bib Hamilton A. J. S., Sarazin C. L., \& Szymkowiak A. E. 1986, ApJ, 300, 698

\bib Hamilton A. J. S., Sarazin C. L., Szymkowiak A. E.,
\& Vartanian M. H. 1985, ApJ, 297, L5

\bib H\"{o}flich P., \& Khokhlov A. 1996, ApJ, 457, 500

\bib Khokhlov A., 1995, ApJ, 449, 695

\bib Kirshner R. P., Winkler P. F., \& Chevalier R. A. 1987, ApJ, 315, L135

\bib Koyama K., Petre R., Gotthelf E. V., Hwang U., Matsuura M., Ozaki M.,
\& Holt S. S. 1995, Nature, 378, 255

\bib Laming J. M.,  Raymond J. C., McLaughlin B. M., \& Blair W. P. 1996,
ApJ, in press

\bib Lennon M. A., Bell K. L., Gilbody H. B., Hughes J. G., Kingston A. E.,
Murray M. J., \& Smith F. J. 1988, J.\ Phys.\ Chem.\ Ref.\ Data, 17, 1285

\bib Livne E. 1993, ApJ, 406, L17

\bib Long K. S., Blair W. P., \& van den Bergh S. 1988, ApJ, 333, 749

\bib Minkowski R. 1966, AJ, 71, 371

\bib Moffet D. A., Goss W. M., \& Reynolds S. P. 1993, AJ, 106, 1566

\bib Morton D. C. 1991, ApJS, 77, 119

\bib Niemeyer J. C., \& Hillebrandt W. 1995, ApJ, 452, 769

\bib Nomoto K., Thielemann F.-K., \& Yokoi K. 1984, ApJ, 286, 644

\bib Raymond J. C., Blair W. P., \& Long K. S. 1995, ApJ, 454, L31

\bib Reilman R. F., \& Manson S. T. 1979, ApJS, 40, 815

\bib Reynolds S. P. 1996, ApJ, 459, L13

\bib Reynolds S. P., \& Gilmore D. M. 1986, AJ, 92, 1138

\bib Reynolds S. P., \& Gilmore D. M. 1993, AJ, 106, 272

\bib Savage B. D., \& Mathis J. S. 1979, ARA\&A, 17, 73

\bib Schaefer B. E. 1996, ApJ, 459, 438

\bib Schweizer F., \& Middleditch J. 1980, ApJ, 241, 1039

\bib Smith R. C., Kirshner R. P., Blair W. P., \& Winkler P. F. 1991,
ApJ, 375, 652

\bib Willingale R., West R. G., Pye J. P., \& Stewart G. C. 1996,
MNRAS, 278, 749

\bib Woosley S. E., \& Weaver T. A. 1987, in
%Mihalas D., Winkler K.-H. A., eds,
IAU Coll.\ 89,
Radiation Hydrodynamics in Stars and Compact Objects,
ed.\ D. Mihalas \& K.-H. A. Winkler
%Springer-Verlag, Berlin,
(Berlin: Springer-Verlag),
91

\bib Wu C.-C., Crenshaw D. M., Fesen R. A., Hamilton A. J. S.,
\& Sarazin C. L. 1993, ApJ, 416, 247 (WCFHS93)

\bib Wu C.-C., Crenshaw D. M., Hamilton A. J. S., Fesen R. A., Leventhal M.,
\& Sarazin C. L. 1996, ApJ Letters, submitted (WCHFLS96)

\bib Wu C.-C., Leventhal M., Sarazin C. L., \& Gull T. R. 1983, ApJ, 269, L5

\end{document}